\documentstyle[11pt,amssymb]{article}

\def\a{\alpha}
\def\b{\beta}
\def\g{\gamma}
\def\d{\delta}
\def\e{\epsilon}
\def\te{{\tilde \varepsilon}}
\def\ve{\varepsilon}

\def\l{\lambda}
\def\lt{{\widehat \lambda}}
\def\m{\mu}
\def\n{\nu}
\def\r{\rho}
\def\rt{{\tilde \rho}}
\def\s{\sigma}
\def\t{\tau}
\def\ti{{\tau^\infty}}
\def\th{\theta}
\def\O{\Omega}
\def\o{\omega}

\def\Gi{{G^\infty}}
\def\Ki{{{\cal K}}}

\def\Hi{{H^\infty}}
\def\Wir{{\cal W}}
\def\L{{\rm L}}
\def\I{{\cal I}}

\def\V{{\cal V}}
\def\Vh{ {\widehat {\cal V}} }
\def\vh{ {\widehat v} }
\def\cL{{\cal L}}
\def\cR{{\cal R}}

\def\K{{K}}
\def\D{{\cal D}}

\def\J{{\cal J}}
\def\Jt{{\widehat {{\cal J}} }}
\def\M{{\cal M}}
\def\S{{\cal S}}

\def\Yh{{\widehat {{\cal Y}} }}

\def\gt{{\tilde g}}
\def\pt{{\tilde p}}

\def\Pt{{\tilde P}}
\def\R{{\bf R }}
\def\C{{\bf C }}
\def\sl{{SL(2,\R)}}
\def\so{{SO(1,1)}}
\def\p{\partial}
\def\Lie#1{{{\rm Lie}}({#1})}

\def\semi{{\Bbb n}}

\def\Ref#1{(\ref{#1})}
\def\baa{\begin{array}}
\def\eaa{\end{array}}
\def\[{\begin{eqnarray}}
\def\]{\end{eqnarray}}
\def\non{\nonumber \\[2mm]}
\def\be{\begin{equation}}
\def\ee{\end{equation}}
\def\lb#1{\label{#1}}
\def\c{\cite}

\def\ft#1#2{{\textstyle {\frac{#1}{#2}} }}
\def\bi{\bibitem}

\begin{document}
\thispagestyle{empty}
\begin{flushright}  hep-th/9608082 \\
                    LPTENS 96/38 \\
                    DESY 96-124 \end{flushright}
\vspace*{0.5cm}
\begin{center}
{\LARGE {Conformal internal symmetry of $2d$ $\sigma$-models
  coupled to gravity and a dilaton}}
\vskip1cm
B.~Julia
\vskip0.2cm
Laboratoire de Physique Th\'eorique CNRS-ENS\\ 
24 rue Lhomond, F-75231 Paris Cedex 05, France
\vskip0.4cm
and 
\vskip0.4cm
H.~Nicolai
\vskip0.2cm
II. Institute for Theoretical Physics, Hamburg University, \\
Luruper Chaussee 149, 22761 Hamburg, Germany \\*
\vskip1.0cm
\begin{minipage}{12cm}\footnotesize
{\bf ABSTRACT:} 
General Relativity reduced to two dimensions possesses
a large group of symmetries that exchange 
classical solutions. The associated Lie algebra is known to
contain the affine Kac-Moody algebra $A_1^{(1)}$ and half
of a real Witt algebra. 
In this paper we exhibit the 
full symmetry under the semi-direct product of $\Lie{A_1^{(1)}}$ by the
Witt algebra $\Lie{\Wir}$. Furthermore we exhibit the corresponding 
hidden gauge symmetries. We show that the theory can be understood
in terms of an infinite dimensional potential space involving all degrees of 
freedom: the dilaton as well as matter and gravitation. 
In the dilaton sector the linear system that extends the previously known
Lax pair has the form of a twisted self-duality constraint that is the
analog of the self-duality constraint arising in
extended supergravities in higher spacetime dimensions. Our results
furnish a group theoretical explanation for the simultaneous
occurrence of two spectral parameters, a constant one ($=y$) and
a variable one ($=t$). They hold for all $2d$ non-linear $\sigma$-models 
that are obtained by dimensional reduction of $G/H$ 
models in three dimensions coupled to pure gravity.
In that case the Lie algebra is 
$\Lie{\Wir \semi G^{(1)}}$; this symmetry acts on a
set of off shell fields (in a fixed gauge) 
and preserves the equations of motion.
\end{minipage}
\end{center}
\newpage

\section{ Introduction}
It has been known for a long time that upon dimensional 
reduction to two dimensions Einstein's theory acquires
a large group of nonabelian symmetry transformations acting
as B\"acklund transformations on the space of classical solutions with
two commuting Killing vectors (stationary axisymmetric or colliding plane
wave solutions) \c{Geroch,GR}. This property turned out to be related to the
integrability of the classical equations of motion, which admit a 
linear system (Lax pair) \c{MBZ} similar to the one governing
the flat space $SO(3)$ nonlinear $\s$-model \c{PL}. After the construction 
of supergravity theories it soon became clear that they share three common
features which one may call silver rules \c{Cremmer, Julia1, Julia3}. 
Hidden symmetries of the solution space (i.e. the on-shell fields) 
form a global non-compact group and the scalar fields 
parameterize a $G/H$ symmetric space with $H$ the maximal compact subgroup 
of $G$ ($H$ is compact in the case of spacelike Killing vectors). 
In even dimension $(2f)$ of space-time the fields that are closed forms 
of degree $f$ constitute a representation of $G$, and some 
combinations  of them involving the scalar fields obey a twisted
self-duality constraint that unifies the Bianchi identities and the equations 
of motion; these precise combinations form a representation of $H$ and the 
twist operator is an 
appropriate invariant of $H$. Thirdly the $H$ stability subgroup can be
gauged and this typically simplifies the transformation rules enormously 
which are somewhat awkward in any fixed gauge. In this paper we shall 
verify the first two rules and partly the third one. 
 
Evidence for the existence
of an underlying affine Kac-Moody Lie algebra of global symmetries,
especially its central charge (that had been overlooked before), 
and for a $\Gi/\Hi$ coset structure with $\Gi$ non-compact
and $\Hi$ its maximal ``compact" subgroup for supergravities
dimensionally reduced to two dimensions 
was found in \c{Julia2}, see also \c{Julia1}\footnote{For a given
group $G$ (but not for the subgroup $H$) we denote by 
$\Gi \equiv G^{(1)}$ the associated full Kac-Moody group (i.e.  
the centrally extended loop group). To be completely precise, we shall 
occasionally append the subscript $t$ or $y$ whenever
the affine parameter to be used is not obvious. Furthermore, for any
Lie group $G$ the associated Lie algebra will
be designated by $\Lie{G}$.}. 
The Lax pair for arbitrary gravitationally coupled symmetric
space ($G/H$) $\s$-models was introduced and studied in detail in \c{BM,Nic} 
for Euclidean resp. Minkowskian signature of the world-sheet
(i.e. the remaining two dimensions).
This work led to the precise implementation of the affine 
Kac-Moody symmetries for these models: the general result is that, 
in two dimensions, the symmetry is enlarged to a group 
$\Gi \equiv G^\infty_y$ which turns out to be the affine extension 
of the (finite dimensional) non-compact Lie
group $G$ that one obtains by first reducing to three dimensions,
and dualizing there any remaining non-scalar matter fields; the 
central charge $c$ of $\Gi$ then acts as a scaling operator 
on the conformal factor of the $2d$ metric.
We note that infinite dimensional symmetries also appear in flat
space $\s$-models \c{Sigma}. Early works on 
symmetries acting on the ``constant scattering parameter" $y$ of 
the affine Kac-Moody algebra are \c{Cosgrove, Witt1, Witt2};
a more recent discussion of the
gravitationally coupled models may be found in \c{JHS}. For a very
different (twistorial) point of view on the Geroch group, 
see \c{WM}. We shall summarize the current knowledge in section 3.

The main result of the present work is that the $2d$ gravitationally 
coupled symmetric space $\s$-models of the type of the models obtained 
by dimensional reduction of (super)gravity theories in dimension at least
equal to three are invariant under the semidirect product of a real form of 
the Witt algebra $\Lie{\Wir}$ and the
affine Kac-Moody algebra $\Lie{\Gi}$. The set of all fields 
including the propagating degrees of freedom as well as 
the conformal factor, the dilaton and a collection of
auxiliary fields take their values in a space that has the same transformation 
properties as the infinite dimensional coset space
\[ \M = \frac{\Wir \, \semi \, \Gi}{\Ki \, \semi \, \Hi}  
\lb{bigcoset} \]
in a fixed triangular (Borel-type) gauge. In the physical interpretation,
$\M$ should be thought of as a ``moduli space of solutions", but there 
is a crucial difference between the dilaton sector (described by
$\Wir/\Ki$) and the matter sector (described by $\Gi/\Hi$): owing
to the $2d$ general covariance of the model, the dilaton field can
in principle be identified with one of the world-sheet coordinates
and therefore does not carry any propagating degrees of freedom. 
We will see, however, that such a gauge choice would obscure the
symmetries of the theory. 

Let us emphasize that our notation is somewhat schematic in that 
the various infinite dimensional
``groups" appearing in \Ref{bigcoset} are considerably more difficult
to come by than their finite dimensional analogs. Nevertheless 
we shall speak of coset fields and try to define the gauge fixing and 
parametrizations involved. We shall explain in section 4 that the  
dilaton sector is simply described by a 
space-time dependent map from $S^2$, the Riemann sphere  
of the complex variable $t$, to the cut $y$ plane; this map 
is even under the involution 
$t \rightarrow 1/t$ for a Lorentzian space-time (in the 
Euclidean case the involution exchanges $t$ and $-1/t$). 
We postpone the consideration of higher genus 
Riemann surfaces to a later paper. 
In section 5 we allow for arbitrary coordinates on the target $S^2$ and 
formally recover the full Witt algebra, which is thereby realized
as a set of non-linear and non-local transformations on the 
dilaton and matter fields; these transformations
are similar to the realization of the Geroch group in the matter sector. The
term matter applies to the propagating gravitons seen from a two dimensional 
perspective as well as to the more general sigma model fields. 
In section 6 we obtain the unifying semi-direct product structure 
and explain the deep analogies between the 
propagating fields and the dilaton itself. The full
set of equations is rewritten in group theoretical form, 
such that the $\Gi$ invariance can be verified.
The $\Hi$ local invariance is fixed in the present treatment that uses a
triangular gauge but should be restored in a further work; the results
presented in section 4 should be viewed as a first step in this direction.
In the last section 7, we will present some concluding remarks and introduce 
a new action principle for the dilaton sector.
But first let us review some results on infinite dimensional Lie algebras.

\section{Mathematical background}
It is not a simple task to extend the theory of finite dimensional Lie groups
and Lie algebras to the infinite dimensional case. Even the definition 
of the Lie algebra itself requires a careful choice of topology 
and completion. The corresponding group 
is an infinite dimensional manifold. 
In particular, the well known statement that any group element 
in the vicinity of the identity can be reached by exponentiation
of a suitable Lie algebra element, familiar from the theory of 
finite dimensional Lie groups, is not general for infinite dimensional 
Lie groups (see e.g. \c{Mil}). 
For example it is not true for the real Lie group 
${\rm Diff}^+(S^1)$ of orientation 
preserving reparametrizations of the circle, 
even if one restricts oneself to real analytic diffeomorphisms. 
One can also prove that the group of $C^\infty$ diffeomorphisms of a manifold 
cannot be given any real analytic structure. 

Following \c{BM}, let us first define the groups $\Gi$ and $\Hi$.
$\Gi$ consists of pairs $(g(y),a)$ where $a$ is a real number and
$g(y)$ is a map $y\rightarrow g(y)$ from the complex
$y$-plane into the complexified group $G$, which is locally
analytic in a neighborhood of the origin $y=0$ (this definition
includes meromorphic maps $g(y)$ as well as maps having more
general singularities than poles). 
In addition, $g(y)$ is subject to the reality 
constraint that, for real $y$, $g(y)$ be an element of 
the relevant real form of the group $G$; e.g. for $G=SL(n,\R )$,
we demand $g(\overline{y})= \overline{g(y)}$ in the defining representation.
The group composition law is given by
\[ \Big( g_1(y), a_1 \Big) * \Big( g_2(y), a_2 \Big) =
   \Big( g_1 g_2 (y), a_1 a_2 e^{\O (g_1, g_2)} \Big)  \lb{group1}  \]
where $\O( g_1 , g_2)$ is the group two-cocycle needed to ``exponentiate"
the central term; the reality constraint on $g(y)$ also
ensures reality of $\O (g_1,g_2)$. For asymptotically flat axisymmetric 
stationary solutions (i.e. the Euclidean case) one needs the 
extra constraint $g(0) = {\bf 1}$ and hence regularity at the origin.

Observe that this definition of $\Gi$ is more general
than the one usually employed in the mathematical literature 
where loop groups are defined as maps from
$S^1$ into the group \c{PS}
(otherwise we would not even get the Schwarzschild solution). 
The complex variable $y$ will be referred to as the
``constant spectral parameter" henceforth. 
Even more generally, one might want to consider higher genus spectral
parameter Riemann surfaces, in which case $y$ should be treated as a local
coordinate (see \c{KM} for examples of solutions for which 
$y$ ``lives" on a hyperelliptic Riemann surface, and \c{KN} for
a treatment of higher genus world-sheets).

The infinite dimensional abstract group $\Hi$
appearing in the denominator of \Ref{bigcoset} is 
parameterized in terms of a second (complex) variable $t$, the
so-called ``variable spectral parameter". We shall discover that these 
qualificatives, constant or variable, are not general, they depend on the 
point of view, namely the choice of a so called active or passive 
point of view to use the terminology of transformation theory. 
For Lorentzian world-sheets, $\Hi$ is defined as the  
maximal compact subgroup of a second $\Gi$ group parametrized in
terms of $t$ (i.e. $G^\infty_t$) 
with respect to the (generalized) 
Cartan-Killing metric on $\Lie{G^\infty_t}$. 
More generally, it is the subgroup invariant under an involution $\ti$ 
defined by $\ti (T_m^a):= \e^m \t (T_{-m}^a)$, where $T_m^a$ are the
affine generators, $\t$ is the involution
defining the finite dimensional symmetric space $G/H$, and 
$\e=1$ ($\e=-1$) for Lorentzian (Euclidean) world-sheets \c{Julia2,BM}.
Henceforth we shall concentrate 
on the Lorentzian case unless specified otherwise.  
We refer to \c{Helgason} for the definition of
maximal compact subgroups $H$ as the invariant set of a suitable 
involution $\t$, i.e. $\t(h)=h$ for all $h\in H$; such involutions exist
for both compact and their dual non-compact symmetric spaces.
Consequently, the group  $G^\infty_t$ consists of maps from 
the complex $t$ plane into $G$ obeying the same reality constraint 
in terms of $t$ as $G^\infty_y$ in terms of $y$ (the compatibility
of these choices will be addressed later when we discuss the relation between 
the real $y$ line and the union of the unit circle and real line of $t$). 
Its subgroup $\Hi$ satisfies the additional constraint
\[ h(t)= \ti \big( h(t) \big) \equiv 
   \t \bigg( h\Big(\frac{1}{t} \Big) \bigg) \lb{h(t)} \]
i.e.
\[h\circ\I = \t\circ  h  \lb{inv}\]
where the inversion exchanging $t$ and $1/t$ has been given a name: $\I$, as 
it is going to be quite important in this paper.
Note that the central charge generator of $\Lie{\Gi}$ is orthogonal 
to $\Lie{\Hi}$. It is null and conjugate to the grading operator for 
the standard invariant bilinear form described for example in \c{Kac}.

The coset space $\Gi/\Hi$, in which  the matter fields  of the
$\s$-model including the conformal factor field take their values, has been 
defined in \c{BM,Nic} to consist of certain matrix functions 
$\Vh(x,t)$ which are acted upon by $g(y)$ and $h(x,t)$ from left 
and right, respectively, with $y$ and $t$ related as in \Ref{y} below.
For Einstein's theory and modulo some technical assumptions 
it can be shown (for Euclidean signature) that $\Gi$ acts transitively on
the manifold of axisymmetric stationary solutions \c{Ernst, BM}.

The coset fields are parametrized in a triangular gauge by demanding 
``regularity" of $\Vh$ in a certain neighborhood of the origin in the
$t$ plane. At the formal Lie algebra level, regularity reduces to requiring 
absence of poles at a given point $t_0$, which in the past has always
been chosen as $t_0=0$. It is clear that the $\Lie{\Hi}$ compensator needed  
to cancel a Laurent series singularity at the origin arising from 
an infinitesimal $\Gi$ always exists. For $\Vh(x,t)$ not satisfying the
regularity condition, one can find an element $h(x,\cdot)\in\Hi$ 
such that $\Vh (x,t)h(x,t)$ is again triangular; this can be
shown by first defining the so-called ($x$-independent) monodromy matrix
$\Vh \ti ( \Vh^{-1})$ and then solving the associated 
Riemann Hilbert problem for it \c{BM}. Uniqueness of the Riemann-Hilbert 
decomposition obtains actually by defining ``triangularity" or 
``regularity" to mean holomorphy of $\Vh(x,t)$ in a neighbourhood $\D_x$
which generically depends on the spacetime position $x$. Furthermore,
$\D_x$ should contain a fundamental domain of the involution $\I$,
such as for instance the open unit disc plus the closed upper semicircle 
connecting the two fixed points $t=1$ and $t=-1$ of 
the involution $\I$. Then
$\D_x \cup \I (\D_x) = \C$, and $\D_x \cap \I (\D_x)$ is an annular
region, on which the Riemann Hilbert problem can be posed. 
One of our main points will be that there is nothing special
about any particular region $\D_x$ or the point $t_0=0 \in \D_x$ 
in this context, because the parameter $t$ is subject to
certain $x$-dependent diffeomorphisms $k: t\rightarrow k(t)$
(to be introduced below). In this sense one might say that 
``the Riemann Hilbert problem should be gauged". 

The definitions of $\Wir$ and $\Ki$ and the associated 
``coset space" $\Wir/\Ki$ are even more subtle.
We would like to think of $\Wir$ as the group of real analytic
reparametrizations $y \rightarrow f(y)$ of the constant spectral
parameter $y$ that is generated by differential operators acting on functions 
of $y$: 
\[ \L_m = -y^{m+1} \frac{\p}{\p y}  \lb{Witty}  \]
obeying the Witt algebra
\[ [\L_m,\L_n] = (m-n)\L_{m+n} \lb{Wira}\]
These analytic reparametrizations do respect the reality constraint on $\Gi$, 
as we have  $f(\overline{y}) = \overline{f(y)}$ by the
Schwarz' reflection principle. The group $\Wir$  we need has not 
been rigorously defined yet and probably  there is no such group. 
One way to describe it would be to think of it as a subsemigroup
of the complexification of ${\rm Diff}^+(S^1)$ which is only a
semigroup \c{Ner,Kir,Sega}. An even weaker notion is that of
a ``local semigroup", but we would prefer a global definition of the 
space of fields, and we may hope to get a pseudogroup, with inverse and 
composition defined on variable  domains of definition. If one could avoid 
the complex domain one could presumably take a group of real differentiable
diffeomorphisms for $\Wir$. 

At the Lie algebra level the situation is easier to handle:
the usual quantum (unitary) Virasoro symmetry (without the central charge)
is precisely the real Lie algebra of ${\rm Diff}^+(S^1)$. Namely,
together with the commutation relations \Ref{Wira}, 
one usually considers elements $\sum_m \a_m \L_m$ 
satisfying the reality conditions
$\a_m=-\overline{\a}_{-m}$ (so $\a_0$ is imaginary). 
By contrast, we here need {\em real} coefficients $\a_m$,
corresponding to a different real form of the complexified Virasoro
algebra (without central charge) it is known in mathematics under the name of
real Witt algebra (the Witt algebra exists over all rings). 

Let us start with the better known and important
 $\Lie{{\rm Diff}^+(S^1)}$, or, 
in other words, with the Lie algebra of real vector fields 
tangent to the unit circle. One cannot exponentiate its full 
complexification. There is an associated
complex semigroup generated by a cone of complex vector fields
on the unit circle, it can be (incompletely) defined by considering all
injective analytic maps from $S^1$ to the unit disk avoiding the origin and
encircling it in the trigonometric sense without critical point. 
The composition
law is simply defined by the composition of these maps when the second one
can be continued analytically all the way to the image of the first. In the 
general situation, however, the product law requires heavier machinery.

${\rm Diff}^+(S^1)$ admits a nice Bruhat type decomposition.
Consider the set $\S$ of holomorphic and univalent (i.e. injective) maps 
$f_0 : D\rightarrow \C$ from the unit disk $D$ into the complex plane
smooth up to the boundary
with $f_0(0)=0$ and $f_0'(0) =1$ (i.e. analytic maps of the form 
$f_0(y)=y\big(1+\sum_{k\geq 1} c_k y^k\big)$); 
univalence amounts to require the coefficients 
$c_k$ to satisfy the Bieberbach-de
Branges inequalities $|c_k| \leq k+1$. By the Riemann mapping 
theorem there exists (up to a rotation) a holomorphic and univalent function
$f_\infty$ from the complement $\C \setminus D$ to $\C$ such that
the interior of $f_\infty \big(\C \setminus D \big)$ 
is contained in the closure of $\C \setminus f_0 (D)$ and
the boundaries of $f_0(D)$ and $f_\infty (\C \setminus D)$ agree,
i.e. $f_0 (S^1) = f_\infty (S^1)$. $f_0$ thus defines an element of
${\rm Diff}^+ (S^1)$  
\[\g := f_0^{-1} \circ f_\infty \big|_{S^1} \lb{gamma} \]
again up to a rotation.
Conversely, given an element
$\g \in {\rm Diff}^+(S^1)$, it is 
possible to define a ``factorization" procedure that
constructs an analytic function $f_0 \in \S$. 
Schematically, one can think of the factorization as the product decomposition:
\[\g =\exp \sum_{n=1}^\infty \Big(\a_n \L_n \Big)
\; \exp \big(\a_0 \L_0 \big)\; 
\exp \sum_{n=1}^\infty \Big(-\overline{\a}_n \L_{-n}\Big).\lb{FAC}\]
This shows clearly that $f_0$ determines $f_\infty$ and $\g$ up to 
constant rotations: in other words it parametrizes ${\rm Diff}^+(S^1)/S^1$.  
The rigorous proof of factorizability involves the uniqueness of the complex 
structure on the Riemann sphere.

Let us now return to the Witt algebra $\Lie{\Wir}$.
We will tentatively define the ``regular" elements of $\Lie{\Wir}$
by the requirement that they should vanish at $y=0$;
they form a triangular subalgebra which is spanned by the operators
$\L_m$ with $m\geq 0$, and which can be exponentiated locally 
to give analytic maps that leave the point $y=0$ fixed. They
correspond to the left factor in \Ref{FAC} suitably modified because we are 
considering a different real form of the Virasoro group. 
Again one sees some element of arbitrariness in the choice of the point $y=0$ 
as the regularity reference point. Our discussion follows 
the discussion of $\Gi$ above. 
While this part of the $\Lie{\Wir}$ invariance can be readily understood, 
a large part of our effort will be devoted to making sense 
of the ``singular" generators $\L_m$  with negative $m\in {\bf Z}$,
corresponding to the right factor in \Ref{FAC}

By analogy with $\Wir$, $\Ki$ should be thought of as the real group of 
transformations of the real variable $t$ 
generated by the operators (for $m\geq 1$)
\[ \K_m := \cL_m - \cL_{-m}          \lb{Wittt1}  \]
where $\cL_m$ are the analogs of the operators \Ref{Witty} on $t$, i.e. 
\[ \cL_m := -t^{m+1} \frac{\p}{\p t} ;  \lb{Wittt2}   \]
we use script letters to distinguish these operators
from the ones introduced in \Ref{Witty} as they act on functions of the 
variable $t$. Hence
$\Lie{\Ki}$ corresponds to the odd vector fields on $S^1$.
It exponentiates to diffeomorphisms of the circle $(t\overline{t}=1)$ that 
satisfy $k(t)=\overline{k(\overline{t})}$. This formula can be extended 
holomorphically to complex values of $t$ as $k(1/t)=1/k(t)$.
As a subgroup of ${\rm Diff}^+ (S^1)$ acting on the unit circle in the
$t$-plane, $\Ki$ is a bona fide group; it is essentially the 
intersection of ${\rm Diff}^+(S^1)$ and ${\rm Diff}^+(\R)$.

$\Ki$ is thus realized as the set of transformations
$t\rightarrow k(t)$ that satisfy
\[ k\Big(\frac{1}{t} \Big) = \frac{1}{k(t)}.     \lb{k(t)}   \]
They preserve (in the semi-direct product structure) the subgroup $\Hi$. 
Indeed \Ref{h(t)} is also satisfied for $\tilde h(t):=(h\circ k)(t)$
if it is satisfied for $h(t)$ provided $k \in \Ki$.
Instead of \Ref{k(t)}, one can equivalently write
\[\I\circ k = k\circ \I.\lb{inv2}\]
The appearance of the involution $\I$ in
\Ref{inv} and \Ref{inv2} introduces a two-to-one ambiguity which we
will encounter time and again: as we will see,
$t$ lives on a double cover of the $y$ plane. Then \Ref{inv}
and \Ref{inv2} relate the actions of $k$ and $h$ 
on the two sheets. 
As a special subgroup of the orientation preserving diffeomorphisms,
$\Ki$ also preserves orientation, whereas $\I$ does not. 
We have thus found an abstract characterization of
$\Lie{\Ki}$: it consists of the elements of
$\Lie{\Wir}$ that are invariant under the involution 
$\ti (\cL_m) := - \cL_{-m} $; it may then be easily seen that 
this definition is compatible with the involution $\ti$
on the Kac Moody algebra introduced in \Ref{h(t)}. One remarks that $\ti$ 
extends to the anti-involution \c{Helgason} defining the real form 
$\Lie{{\rm Diff}^+(S^1)}$ from its complexification as one expects from the 
last remark of the previous paragraph. The relation between $\Wir$ and 
$\Lie{{\rm Diff}^+(S^1)}$ is precisely the relation between 
a non-compact Lie algebra and its dual compact real form in E. Cartan's theory
of Riemannian symmetric spaces.
Here however, and in contrast to the Kac-Moody algebra $\Lie{\Gi}$, 
there exists no symmetric $\Wir$-invariant bilinear form on $\Lie{\Wir}$
\c{Kac}, and consequently no definition of
the elements of $\Lie{\Ki}$ in terms of their 
norm\footnote{A similar phenomenon takes place for indefinite (simple) 
Kac Moody algebras even though these do possess an invariant bilinear form. 
As an example consider the Kac Moody algebra constructed from the
Chevalley generators $(e_i,f_i,h_i)$ with symmetric
Cartan matrix $A_{ij}$ \c{Kac}. 
The Chevalley involution is 
$$ \th (e_i) = -f_i \;\;\; , \;\;\; \th (h_i) = - h_i$$
and the norm of any element of the Kac Moody algebra can be 
computed from the standard bilinear form
$$
\langle e_i| f_j \rangle = \delta_{ij} \;\;\; , \;\;\;
\langle h_i| h_j \rangle = A_{ij}
$$
Clearly, for positive definite
or semi-definite $A_{ij}$, the set of negative norm generators
$e_i - f_i$ and their multiple commutators
coincides with the set of $\th$-invariant ones. This set does
not include any elements from the Cartan subalgebra and
closes into a subalgebra of the full Kac Moody algebra, which for 
$\Gi= SL(n,\R)^\infty$ just corresponds to the subgroup $SO(n)^\infty$ 
defined by \Ref{h(t)} (so in this case we have $\ti = \th$). 
For indefinite $A_{ij}$, on the other hand,
there exists at least one 
element $h= \sum_i n_i h_i$ of the Cartan subalgebra with
$\sum_j A_{ij} n_j = \kappa n_i$ and $\kappa < 0$ such that
$$ \langle h | h \rangle = \sum_{i,j} n_i A_{ij} n_j = 
     \kappa \sum_i n_i^2 < 0    $$
But then
$$ [ h, e_i - f_i ] = \kappa (e_i + f_i)     $$
and, in contrast to the $\th$-invariant elements, the negative norm
elements of the Kac Moody algebra no longer form a subalgebra
since $e_i + f_i$ has positive norm.
This may be relevant for the possible emergence of hyperbolic Kac Moody 
algebras in the further reduction of Einstein's theory.}. 
Yet another property of $\Ki$ is familiar  
from string theory: it is the largest anomaly free subalgebra 
of the Virasoro algebra \c{Gross}. This is analogous to the
fact that $\Hi$ does not contain the central charge of $\Gi$.
Note the general explicit formula
\[ \exp \big( \a \K_m \big) (t) =
 \bigg( \frac{ t^m \cosh (m\a)+\sinh (m\a)}{t^m \sinh (m\a)+ \cosh (m\a)}
 \bigg)^{\frac{1}{m}}      \lb{expKm}   \]

Again requiring regularity  at $t=0$ seems analogous to a choice
of triangular gauge and is precisely that at the formal Lie algebra level. 
However, a precise definition that would permit the discussion of 
singularities away from $t=0$ or, at the group level, the 
factorization theorem analogous to \Ref{FAC} corresponding to the 
Riemann-Hilbert problem is more problematic: the latter should be
derived for the real form $\Lie{\Wir_t}$, 
furthermore regularity should mean univalence on a disc or 
maybe on a fundamental domain of the involution $\I$. 
We hope to return to these issues in a future paper dealing with gauge 
restoration.  There we would need
two types of factorization results: the analog of \Ref{FAC},
familiar in the theory of dressing transformations, 
that is similar to a Bruhat 
decomposition and the more usual one for (super)gravities which is of the 
Iwasawa type. In this paper we do use a formal form of the latter to
compute the compensating $\Hi$ and $\Ki$ gauge transformations.  
We shall ignore the fact that 
a (unique) Iwasawa decomposition of a diffeomorphism $\g(t)$ does not
always exist for the Lorentzian world-sheets (see our comment in section 4).
In fact the set of fields will turn out not to be exactly a coset space but 
will possess the same symmetries, so the decomposition and gauge fixing
problems take a slightly different form.

Let us now add some new mathematical observations for later use.
In the dilaton sector of the theory it will be sufficient to 
consider even functions under the involution $\I$.
Implicitly, these are functions of the variable 
\[ q(t):=\frac{2t}{1+t^2}.\lb{q} \] 
For small $t$ we have $q \sim 2t$ and near infinity $q \sim 2/t$. We leave it 
as an exercise to show that the real Lie algebra of $\Ki_t$ acts as the set of 
real regular vector fields on the completed ($q$) real line minus the open 
interval $]-1, 1[$, that vanish at the two end points. In other words one can 
map $\Ki_t$ to the (orientation preserving) diffeomorphism group of a closed 
interval of $1/q$. 
In particular one obtains
\[ \K_1 := (1-t^2) \frac{\p}{\p t} 
  = 2 (1-q^2) \frac{\p}{\p q}. \lb{acc} \]
The fact that these are the same operators (up to a factor)
will play a role in section 5. More generally, $q$-vector fields
are ``pulled back" according to the formula
\[ t_{\ast} \bigg( -q^{n+1} \frac{\p}{\p q} \bigg) =
 - \frac{(2t)^n (1+t^2)^{1-n}}{1-t^2} \,  t\frac{\p}{\p t}  \lb{t(q)}  \]
so all the Witt generators for $q$ are mapped into $t$ vector fields
having possibly simple poles at $t=\pm 1$. Admitting such poles, 
we can therefore embed the Witt algebra $\Lie{\Wir_q}$ into some singular
extension of $\Lie{\Ki_t}$; at the formal level, this is
only possible because $\Lie{\Ki}$ is infinite dimensional, more precisely the
two Lie algebras to be considered have different behaviours at 
$t=\pm1$\footnote{In fact,
there exist infinitely many such embeddings: one needs only
replace $q(t)$ in \Ref{q} by other functions symmetric
with respect to $\I$.}. It will be important later that the image of 
$\Lie{\Ki_q}$ corresponds to regular elements of $\Lie{\Ki_t}$ at $t=\pm1$. 

We can now be a little bit more specific about the contents of this paper.
In section 4 the M\"obius subgroup $\sl \subset \Wir$  (generated by
$\L_{\pm 1} $ and $\L_0$) and
$\so \subset \Ki_q$ will play an important role;
they correspond to bijections of the Riemann sphere $S^2$.
$\so$ consists of the special M\"obius transformations generated by $\K_1$.
$\so$ is noncompact for Minkowskian signature but,
for Euclidean world-sheets, $\so$ would have to be replaced by the
compact group $SO(2)$. In the Euclidean signature case
$t$ is pure imaginary and is usually written as $it$ for real $t$.
In section 5, we will give a precise definition of the 
``coset space" $\Wir/\Ki$ in terms of (equivalence classes of) maps 
$Y: t\rightarrow Y(t)$, which are acted upon 
by $\Wir$ ($\Lie{\Wir}$ being the part we control) and $\Ki$
from left and right, respectively. As usual any element of this coset should
be viewed as some kind of infinite dimensional vielbein. 
In analogy with the coset fields in $\Gi/\Hi$ of the variable $t$, 
which describe the
matter sector, we shall refer to $\Wir/\Ki$ as the dilatonic sector, 
because the coset space is parameterized by the
dilaton field $\r$ and certain dual potentials; it differs from the matter 
coset space, however, in that the dilaton, 
whose properties it describes, is not an independent 
physical degree of freedom. Actually as we shall see, 
the equations of motion couple 
it to the conformal structure and it reacts to it.
The dilaton dynamics is generated by a linear system very similar
to the one governing the matter sector (see \Ref{LS-t} and \Ref{LS-t2} below), 
with the only difference that $\r$ is a free field and therefore, on-shell, 
all its higher order dual potentials differ from $\r$ or its
dual field $\rt$ only by integration constants (we are ignoring
possible subtleties with dualization for topologically
non-trivial worldsheets). As we will demonstrate in sections 4 and 5 
the dilaton sector relates the two kinds of  spectral parameters $t$ and $y$.
In section 6 we shall first derive the transformation laws of a set of 
off-shell fields (including auxiliary fields) in a triangular Borel gauge 
by using in intermediate stages the equations of motion. 
We shall then turn to the invariance of the extended set of equations of motion,
the extended linear system, and rewrite the conserved current of global
symmetry.

\section {The state of the art} 
We shall make our best effort to keep this paper self-contained but in 
the remainder we will adhere as far as possible to the conventions and 
notations of \c{Nic}. Here we are 
mostly concerned with the Lorentzian case (corresponding to
colliding plane wave solutions) and merely indicate the 
modifications needed for Euclidean worldsheet metrics
(corresponding to stationary axisymmetric solutions) whenever appropriate;
see \c{BM} for further details.
In addition to the matter fields and the dilaton to be described below
the model contains the gravitational worldsheet degrees of freedom
consisting of
the conformal factor $\l$ and the unimodular part of the worldsheet metric.
The latter is purely topological and can always be gauged away
locally; we will therefore ignore it throughout most of this paper.
This means that we will work in the conformal gauge, but of course
all of our results remain valid in a generally covariant setup
(cf. section 7). We will find it occasionally 
useful to employ lightcone (conformal) coordinates $x^+$ and $x^-$.

As for the matter degrees of freedom, we shall assume  that
the scalar fields live on a finite dimensional (non-compact) symmetric space 
$G/H$
and that the local $H$ invariance has been restored already. If one is doing 
dimensional reduction from three dimensions this can be 
done in three dimensions. They are thus described by a
matrix $\V (x) \in G$ subject to the transformations 
\[ \V (x) \rightarrow g \V (x) h(x)  \lb{sigmatrafo}   \] 
with  arbitrary $g \in G$ and $h(x) \in H$.  
The dynamics is defined via the Lie algebra decomposition
\[ \V^{-1}\p_\m \V = Q_\m + P_\m   \lb{VdV}     \]
with $Q_\m \in \Lie{H}$  and $P_\m \in \Lie{H}^\perp$ (the orthogonal  
complement of $\Lie{H}$ in $\Lie{G}$, which is assumed to be semi-simple,
with respect to the Cartan-Killing metric). In addition to the matter 
fields, the model contains a dilaton field $\r$, which is just the 
determinant of the internal metric of the Kaluza-Klein reduction \c{Julia2,BM},
or equivalently the 33-component of the dreibein in the dimensional
reduction from three dimensions \c{Nic}. The dilaton field satisfies
a free field equation, so we have
\[ \r (x) = \r_+ (x^+) + \r_- (x^-)  \lb{rho}   \]
with left- and right-moving solutions $\r_+(x^+)$ and $\r_-(x^-)$. We can
therefore introduce the conjugate harmonic function 
(``axion") by
\[ \rt (x) := \r_+(x^+) - \r_- (x^-) \lb{rhotilde}   \]
With the flat metric $\eta_{00}=-\eta_{11}=1$ and the convention $\ve_{01}=1$, 
we thus have 
\[ \p_\m \r = - \ve_{\m \n} \p^\n \rt. \lb{harmr} \]
For Euclidean world-sheet metrics $\r$ would be the imaginary part
of an analytic function, and the associated real part $z$ the
analog of $\rt$. As is well known \c{McCallum},
one can then identify the complex field $\xi:=z+i\r$ with the worldsheet 
coordinates by means of a conformal coordinate transformation
such that $\r \geq 0$ becomes the radial coordinate
and $z \in \R$ parametrizes the symmetry axis (so-called Weyl canonical 
coordinates). For colliding plane wave solutions (Lorentzian world-sheets),
the appropriate gauge choice is $\r(x^+,x^-)= 1- (x^+)^2 - (x^-)^2$
(see e.g. \c{Nic}). However, we do not want to rely on such gauge choices 
here as they are local only and hide the symmetries we are about to expose.

The matter field equations of motion read 
\[ D^\m (\r P_\m) = 0 ,  \lb{equP}    \]
where $D_\m$ is the $H$-covariant derivative and indices are 
pulled up and down with the flat Minkowski metric
in the conformal gauge. There are two
(compatible) first order equations for the conformal factor, viz. 
\[ \p_+ \r \, \p_+ \s = \ft12 \r \, {\rm Tr} (P_+ P_+)  \non
   \p_- \r \, \p_- \s = \ft12 \r \, {\rm Tr} (P_- P_-)  \lb{equconf}     \]
where $\s$ is defined by
\[ \s =\log \lt :=\log \l -\ft12 \log \big( \p_+\r \, \p_-\r \big) 
             \lb{loglt}     \]
The reason for the $\r$-dependent modification is that $\s$ 
is a bona fide scalar field in contrast to $\log \l$, which 
transforms as the logarithm of a density\footnote{Although it would seem
that the definition \Ref{loglt} is only possible in the conformal gauge,
a covariant definition can be given with the aid of the calculus
of Beltrami differentials \c{Nic2}.}. It can be
seen that if $\r=const$, \Ref{equconf} would imply $P_\m =0$ by the positivity
of the Cartan-Killing metric on $\Lie{H}^\perp$. This is just the
mathematical expression of the fact that the model does not possess
a non-trivial flat space limit.

The matter field equation \Ref{equP} for $\Vh (x,t)$ as well as the usual 
integrability
conditions following from \Ref{VdV} can be recovered as the compatibility
conditions of the linear system (Lax pair)
\[  \Vh^{-1} D_\m \Vh 
   = \frac{1+t^2}{1-t^2} P_\m +
              \frac{2t}{1-t^2} \Pt_\m  ,    \lb{LS-matter} \]
where $\Pt_\m := \ve_{\m \n} P^\n$, $D_\m \Vh := \p_\m \Vh - \Vh Q_\m$,
and $t$ is the
$x$-dependent spectral parameter\footnote{The linear
system for Euclidean signature in the conventions of \c{BM} 
is obtained by replacing $t\rightarrow it$ and 
$\r_\pm \rightarrow \r \pm iz$ in all formulas, where $(\r,z)$ are 
the standard Weyl canonical coordinates.}  
(on which $\p_\m \equiv \p_\m\big|_y$ acts).
Until the end of section 4 partial 
derivatives with respect to the worldsheet coordinates will 
thus always mean derivatives at constant $y$. The matrix
$\Vh = \Vh (x,t)$ extends the original scalar field matrix $\V (x)=\Vh (x,0)$.
When expanded in $t$, it gives rise to 
an infinite number of new fields (dual potentials); these
are not independent, however, but severely 
constrained by \Ref{LS-matter} to depend on $\V$ in a non-linear
and non-local manner. \Ref{LS-matter} is invariant under the product of
(space) parity and the exchange of $t$ with $-t$ as well as under 
$\ti=\t\circ\I$. 
As we are mainly interested in the group theoretical
structure underlying the above linear system and its
generalizations, we will not dwell on the important applications
of \Ref{LS-matter} for the construction
and systematic investigation of solutions of Einstein's equations here.
We would, however, like to
emphasize some unusual and so far unexplained features of the above
linear system, and to reformulate them in a suggestive way. 

First of all, the consistency of \Ref{LS-matter} requires that the spectral 
parameter $t$ itself be 
subject to a very similar system of linear differential equations to 
\[ t^{-1}\p_\m t = \frac{1+t^2}{1-t^2} \r^{-1} \p_\m \r
                + \frac{2t}{1-t^2} \r^{-1} {\tilde \p}_\m \r ,  \lb{LS-t}   \]
that relates $\r$ and $t$ as a B\" acklund duality.
By invariance under $t\rightarrow 1/t$, this equation admits two
solutions $t_\pm (x,w)$ which are explicitly given by
\[ t_+(x,w) := \frac{1-u}{1+u} 
 =\frac{\sqrt{w+2\r_+}-\sqrt{w-2\r_-}}{\sqrt{w+2\r_+}+\sqrt{w-2\r_-}} 
 =  \frac{1}{t_-(x,w)}               \lb{t}   \]
(we will often omit the label $\pm$ and write $t\equiv t_+$), or 
\[ u  := \frac{1-t}{1+t} = 
   \sqrt{ \frac{w - 2\r_-}{w + 2\r_+} }  
          \lb{u}   \]
where $w$ is the constant of integration for the first order 
differential equation \Ref{LS-t}.
The parameter $u$ becomes prominent when one rewrites
the linear systems as respectively
\[  \Vh^{-1} D_\pm \Vh = u^{\pm 1} P_\pm \lb{chi-matter}\]
and
\[t^{-1}\p_\pm t = u^{\pm 1} \r^{-1} \p_\pm \r \lb{chi-t}\]
A very peculiar group theoretic property is the fact that 
the r.h.s. of \Ref{LS-matter}
is an element of the subalgebra $\Lie{\Hi}$; another observation
is the common, very special and singular $t$-dependence of \Ref{LS-matter} and
\Ref{LS-t}, i.e. the absence of higher powers of 
$u$ and $u^{-1}$ in \Ref{chi-matter} and \Ref{chi-t}.
The exchange of $u$ and $1/u$ (i.e. 
the exchange of $t$ with $-t$) is a symmetry of the equations 
of motion only when combined with parity, so we note here 
that worldsheet parity is 
superficially broken by the linear system.

For group theoretical considerations 
it is often more convenient to work with the inverse of $w$: 
\[ y:= \frac{1}{w}&=& \frac{2t}{\r (1+t^2) -2t\rt} = \non
   &=& \left\{ \begin{array}{ll}  
            2\r^{-1}t + 4\rt \r^{-2}t^2 + O(t^3)  
\;\;\;\; &{\rm as} \;\; t\sim 0 \\
            2\r^{-1}t^{-1} + 4\rt \r^{-2}t^{-2} + O(t^{-3})
                      & {\rm as} \;\;   t\sim \infty
            \end{array}  \right.       \lb{y}     \]
where we have used \Ref{t}. Equivalently
\[ \frac{1}{y}:= w = \ft12 \r \Big( t+\frac{1}{t}\Big) -\rt.  \lb{w}   \]
As already indicated in the introduction,
we will refer to $t$ (or $u$) and $y$ (or $w$) as the 
``variable" and ``constant" spectral parameters, respectively.
$u$ is a convenient scattering parameter to utilize in lieu of $t$ in  
chiral expressions.
One sees from \Ref{t} that $t$ is a coordinate on a double cover of the
complex $y$-plane with an $x$-dependent cut extending from $-(2\r_+)^{-1}$ to
$(2\r_-)^{-1}$. The end points of the cut correspond to the fixed points
$t=\pm 1$ of the involution $\I$. We shall later  be led to abandon
the traditional point of view according 
to which $y$ or $w$ are constant (i.e. independent of $x$, to be 
completely precise) and $t$ is a divalent function of them, and
instead view the function $y(t,x)$ together with its generalizations
as the basic object.
$\I$ preserves $y$ and exchanges the two sheets with both $t=0$ 
and $t=\infty$ corresponding to $y=0$ (or $w=\infty$). Note also that $\I$
exchanges $u$ and $-u$. 

Secondly, the matter field equations, which for dimensionally reduced pure 
Einstein gravity are those of the curved space sigma model for the symmetric 
space $G/H=\sl/SO(2)$ with a dilaton, are not manifestly symmetric under the 
affine 
extension $G^{(1)}$ of the Ehlers group $G=\sl$. As pointed out above, 
one must first extend the physical scalar fields
to a highly redundant set of dual potentials 
through the replacement of $\V$ by $\Vh$ so as to be able to express
the action of the Geroch group in a simple manner. On top of this 
one may or may not
restore the local gauge invariance under the group $\Hi$ by adding some 
more (gauge) off-shell degrees of freedom. If one were to do so,
the full symmetry group of the matter system would become the product
of a global group $\Gi$ and a local (gauge) group $\Hi$, such that
\[ \Vh(x,t) \longrightarrow g(y) \Vh(x,t) h(x,t) 
     ,   \lb{GiHionV}   \]
generalizing \Ref{sigmatrafo}. 
As we have mentioned in the previous section, the local gauge invariance under
$\Hi$ is usually fixed and $\Vh$ is in a
generalized ``triangular" (Borel or ``regular") gauge, where $\Vh$ is analytic
as a function of $t$ in a well chosen disc around $t=0$ so that
$\Vh (x,t)= \V (x) + O(t)$ (the opposite gauge choice would correspond 
to analyticity around  $t=\infty$). This gauge is very useful in extracting
the physical fields (and hence the space-time metric solving
Einstein's equations) once a new solution of \Ref{LS-matter} has
been obtained. Actually this gauge is even mandatory 
in the usual presentation of the linear 
system because the (original) matter field $\V$ 
is nothing but $\Vh(x,t=0)$.

However, working in a fixed gauge has the drawback that the
Geroch group is realized non-linearly and non-locally, and
its explicit realization is therefore somewhat cumbersome.
Let us repeat that typical field theories lead to a gauge fixed
linear system at first and it is the physicist's task to restore
the hidden simplifying $\Hi$ gauge symmetry.
Now while the infinitesimal action of the Geroch group 
on $\Vh$ after gauge invariance 
restoration is simply given by $\d \Vh = \d g \, \Vh$,
we must in the triangular gauge allow for a uniquely defined compensating 
$\Hi$ rotation that maintains the triangular gauge for $\Vh$, whose 
total variation is given by\footnote{We use $\Delta$
for the total variation including compensating transformations.}:
\[ \Delta \Vh (x,t) = \d g (y) \, \Vh (x,t) 
   + \Vh (x,t) \, \d h \big(\Vh(x,t), \d g (y)\big)  \lb{Geroch1}  \]
Specifically, if $\d g(y) = y^m \d g_m$ with $\d g_m \in \Lie{G}$
for strictly positive $m$, we can put $\d h_m =0$. By the invariance 
of \Ref{LS-matter} under such transformations, the physical fields 
do not vary, and the only effect of the transformation
is to shift the higher order dual potentials in $\Vh$ by 
$\Lie{G}$-valued constants. Therefore this half of $\Gi$ simply
takes care of the infinitely many integration constants that are
left undetermined in the higher order dual potentials. 
For $m=0$, both $\d g_0$ and $\d h_0$ belong to $G\subset \Gi$ 
and $H\subset \Hi$, respectively, and we recover the special symmetries
already present in the original $G/H$ $\s$-model. Finally, 
for negative powers of $y$, the first term on the r.h.s. 
of \Ref{Geroch1} introduces singularities; consequently
one must express $y$ in terms of
$t$ according to \Ref{y} and remove the poles in $t$ by suitable
$\Hi$ transformations. Only the latter are visible on the physical
fields $Q_\m$ and $P_\m$, whose variation can be determined from
\[ \Delta_m \bigg( \Vh^{-1} \p_\m \Vh \bigg) = \p_\m \d h_m +
   \Big[ \Vh^{-1} \p_\m \Vh \, , \, \d h_m \Big]   \lb{LnegV1} \]   
together with 
\[ \V^{-1} \d \V := \Vh^{-1} \Delta \Vh \Big|_{t=0} \lb{dt0}\]
It is important here that the the on-shell transformations
\Ref{Geroch1} preserve the $t$-dependence on the r.h.s.
of \Ref{LS-matter}; this can be verified either by direct
computation or by noting that, for the new triangular $\Vh$
obtained after performing the variation \Ref{Geroch1}, 
the poles at $t=\pm 1$ arise only from the action of $\p_\m$ on $t$
and therefore higher order poles at $t=\pm 1$ are absent.
It has remained a mystery until now why one encounters
a variable spectral parameter and why two parameters are needed 
to realize the Geroch group. 

Third, the realisation of an affine Kac-Moody symmetry
at the classical level suggests the presence of an 
associated Witt symmetry.
The first hints of this extra symmetry
appeared in \c{Cosgrove}, where its M\"obius subgroup was exhibited;
these results were extended to one half of the Witt algebra in
\c{Witt1}. In \c{Witt2} it was realized that this algebra should correspond 
to infinitesimal diffeomorphisms acting on the constant spectral
parameter $y$, which are generated by the operators \Ref{Witty}, 
in terms of which the standard commutation relations of 
the semidirect product $\Wir \, \semi \, \Gi$ can be easily verified
(instead of \Ref{Witty} we could alternatively have 
used ${\tilde {\rm L}}_m \equiv -w^{m+1} \frac{\p}{\p w} = - \L_{-m}$).
The complications introduced by Weyl coordinates have been partly removed,
but the main difficulty, which has not been satisfactorily 
resolved (or even addressed)
so far, lies in making sense of \Ref{Witty} for the full
range of integers $m$, i.e. both for the ``regular" values $m\geq 0$
leaving the point $y=0$ fixed, and the ``singular" values $m\leq -1$,
for which $y=0$ is not left invariant and even may be sent to infinity, 
and in implementing all these 
transformations on all the fields. As we will see this requires
suitable compensating transformations just as for the Geroch
group itself, cf. \Ref{Geroch1}. We shall in particular recover
previous results in a suitable triangular $\Ki$ gauge. 
We shall again see that the special point $t_0=0$ is not preferred
and could in principle be replaced by any other point in the
domain $\D_x$ subject to the conditions stated before. 
A further question which has remained open is whether there exists
a non-trivial central term extending the Witt algebra
to a full Virasoro algebra. Finally the possibility of 
implementing these symmetries off shell has not yet been considered.

\section{A free field harmonic coset for the dilaton sector.}
To illustrate the basic mechanism, we will first 
show how to reinterpret the equations of motion for $\r$ and $\rt$
as a ``twisted" self-duality constraint for an $\sl/\so$ nonlinear $\s$-model.
Although at first sight this would seem to be a rather 
roundabout method to derive a set of free field equations, we will
present evidence below that it is actually part of a much more general
structure. Most importantly, this non-linear $\sigma$-model structure 
will explain the group theoretical significance
of the two spectral parameters $w$ and $t$, as well as the quadratic equation
\Ref{w} relating them. $\r$ is the only field 
not to have been endowed with a group 
theoretical interpretation yet and eq. \Ref{chi-t} suggests some chiral 
structure with both $\r_\pm$ appearing independently, where 
we define $\r_\pm := \ft12 (\r \pm \rt )$ irrespective of whether
$\r$ and $\rt$ are free fields or not. If we consider $\r$ and $\rt$ 
as independent variables we go de facto off shell.
The main idea then is to identify the fields $\r$ and
$\rt$ with the coordinates on the coset space $\sl/\so$; for this purpose
let us define the $\sl$ matrix
\[  v = \frac{1}{\sqrt{\r}} \pmatrix{\r & \rt \cr
                                       0 & 1 \cr}  \lb{v}    \]
This particular parametrization is suggested by 
the similar parametrization commonly used to realize the Ehlers group $\sl$
on the coset space $\sl/SO(2)$ and by the form of the transformations of
$\r$ and $\rt$ under the $\sl$ (M\"obius) subgroup  of $\Wir_y$ 
which coincide with 
the variations of $\r$ and $\rt$ induced by $\L_0,\L_{\pm 1}$ \c{Witt1}. 
All these transformations are chiral in the sense 
that they do not mix $\r_+$ with $\r_-$.

\Ref{v} could also have been arrived at from 
\Ref{y} that can be rewritten as
\[ y = \frac{q}{-\rt q+\r} \lb{yq}\]
with $q$ as in \Ref{q}. 
This formula will play an important role in the rest
of this paper, so it is important to understand it precisely from the group
theoretical point of view.
The M\"obius subgroup $\sl$ of $\Wir$ made its entry surreptitiously
already into eqs.~\Ref{u} as well as in \Ref{yq},
so let us collect here some  relevant formulas. 
The transformation rule 
\[ y(q)=(aq+b)/(cq+d) \lb{proj}\]
 has a group
theoretical interpretation as the action of $\sl$ on its coset space
$\sl/T_+$ where $T_+$ is the upper triangular subgroup (of dimension two).
Instead of the conventional $\sl $ matrix representation 
\[ \pmatrix{a & b \cr c & d }\]
we shall use the  equivalent contragredient one
\[ \pmatrix{ d & -c \cr -b & a} \]
in order to implement \Ref{proj}.
Concretely,
\[ s_{-y}:= \pmatrix{1 & 0 \cr -y & 1 \cr} = \pmatrix{d & -c \cr -b & a \cr}
\pmatrix{1 & 0 \cr -q & 1 \cr}\pmatrix{(d+cq)^{-1} & c \cr 0 & (d+cq) \cr}.
\lb{TrM}\]
Eq. \Ref{TrM} is one instance of a coset parametrization  in a triangular
gauge and one may recognize on the right the compensating $T_+$ (i.e. upper 
triangular) transformation needed to stay in
the strictly lower triangular gauge parametrization of the coset. This is
typical of a left global group action on a coset space. We shall encounter
the exact analogue of this but with $H$
the maximal "compact" subgroup of $G$ replacing the subgroup $T_+$ and the
coset coordinate will
typically be a field. Note that $u^2$ and $w$ are also homographically related.
In the special case of the map \Ref{yq} we extract from \Ref{TrM} the equation
 \[ s_{-y} = v\, s_{-q} \, T(v,q)   \lb{Mob2}   \]
with $v$ defined in \Ref{v} and
 the appropriate upper triangular matrix $T(v,q)$. The mapping and the
operator $v$ correspond in the dilaton sector to the matter operator
$\Vh$ and will have to be generalized
if one wants to push the analogy further and reexpress \Ref{LS-t}
exactly as \Ref{LS-matter}.
 
Two comments are required here, firstly we have slightly abusively 
identified the matrix group $\sl$ appearing in \Ref{Mob2} and its two to one 
image: the group of projective (or homographic) maps of the complex variable q 
as it would appear in 
\[ y(t) = \big(v \circ q \big)(t); \lb{coprod} \]
 this is not serious as the positivity
of $\r$ will provide a well defined choice of sign. The second point is that 
if we insist on the parametrization \Ref{v} we must choose the (contragredient)
 group action \Ref{TrM} in order to reproduce \Ref{yq} and we must use strictly
lower triangular representatives of the coset space by 
the upper triangular gauge subgroup. Alternatively taking the contragredient 
parametrization of $v$ would allow a more conventional realisation of the
Moebius group.

Now in analogy with \Ref{sigmatrafo}, 
the groups $\sl$ and $\so$ act on $v$ itself
according to $v \rightarrow s v h$ with $s\in \sl$ and
$h\in \so$. The triangular form of $v$ in \Ref{v} is due to the gauge 
fixing of the $\so$ gauge symmetry. It can be viewed as a simple example of
an Iwasawa decomposition, but we caution the reader that the latter
is guaranteed to exist globally only when the (gauge) subgroup is compact. In 
fact, one can convince oneself by explicit computation that not every 
$\sl$ matrix can be brought into triangular gauge by an $\so$ rotation. 
We may note that here the Euclidean theory is 
again better behaved than the Minkowskian one.

The global $\sl$ transformation preserves the triangular form of $v$ if 
accompanied by a compensating $\so$ gauge transformation.
The three generators of $\sl$ are linearly represented by 
\[ \L_{1} = \pmatrix{0&1\cr 0&0 \cr}    \;\;\; , \;\;\;
   \L_0 = \pmatrix{-\ft12 &0 \cr 0& \ft12 \cr}  \;\;\; , \;\;\;
   \L_{-1} = \pmatrix{0&0 \cr -1 &0 \cr} ;        \lb{Moebius1} \]
the $\so$ subgroup is generated by $\L_1-\L_{-1}$, we keep the same
notation as for the action of $\Wir$ on functions without much hesitation. 
On the one hand
no compensating rotation is required for $\L_1$ and $\L_0$, 
which transform the fields $\r$ and $\rt$ according
to $\d_1 \r =0 \, , \, \d_1 \rt = - 1$ 
and $\d_0 \r = \r \, , \, \d_0 \rt = \rt$, 
respectively. The variations $\d_n$ have been defined so as to satisfy the
same commutation relations as the $\L_n$. Note that $\L_{1}$ shifts the 
``dual potential" $\rt$ by a constant, by \Ref{w} this should be related to a 
constant shift of $w$; this feature will
recur for the higher order generators $\L_{m}$.

On the other hand, let us consider the transformation 
$s = s_{y_{-1}}:=\exp(-y_{-1}\L_{-1})$ multiplying  $v$ on the left;
(it corresponds to a constant shift of $y$ by $-y_{-1}$ by virtue of  
the isomorphism of the linear 
$\sl$ with the Moebius subgroup of \Ref{Witty}).
Acting  on $v$, $s$ does require a compensating $h$ rotation to 
restore the triangular gauge, viz.
\[  v (\r,\rt) \longrightarrow v_{y{-1}} \equiv v (\r' , \rt')
= \pmatrix{1&0\cr y_{-1}&1\cr} \, v(\r,\rt) \, 
    \pmatrix{ \cosh \! \th & \sinh \! \th \cr
              \sinh \! \th & \cosh \! \th \cr}    \lb{L1trafo}   \]
Demanding that the new matrix $v_{y_{-1}}$  be triangular, we find
\[ \r \coth \! \th + \rt = -y^{-1} \equiv -w .  \lb{wq}   \]

Note that upon substituting $y=y_{-1}$ and 
\Ref{trigo} we recover precisely equation \Ref{w} provided we define:
 $\th (x)$ by $u:= e^\th$ ($\th$ obeys
a free field equation just like $\r$ and $\rt$). Let us
record the elementary formulas
\[ \cosh \! \th =\frac{1+t^2}{1-t^2}\;\;
, \;\; \sinh \! \, \th=\frac{-2t}{1-t^2}\;\; , \;\;
\tanh \! \th = -q(t). \lb{trigo}\]
This ``coincidence" will be explained shortly. It was very suggestive that, 
as for $\Vh$ in the matter sector, \Ref{L1trafo} shows that $v$ has 
a $y$ side on the left and a $t$ (or at least $q(t)$) side on the right, and
it led us to the above definitions. 

But for the time being we should return to the case of a general constant 
$\L_{-1}$ parameter $y_{-1}$,
then the analog of \Ref{L1trafo} determines both the compensating 
transformation and the 
variation of the fields. Quite generally let us insist that this does not
require any use of the equations of motion provided that infinitesimally 
there is a unique factorization theorem of the Iwasawa type at hand.
Writing $\r (x;y_{-1})$ instead of $ \r'(x)$ and  $\rt (x;y_{-1})$
instead of $ \rt' (x)$, and 
suppressing the dependence on the worldsheet coordinate $x$, we find
\[ \r(y_{-1}) &=& \frac{\r}{(1+2y_{-1} \r_+)(1-2y_{-1}\r_-)} 
    = \r + y_{-1}\big[-2 \r \rt\big] + O(y_{-1}^2) \non
  \rt(y_{-1}) &=&
  \frac{\rt + y_{-1}(\rt^2 - \r^2)}{(1+2y_{-1}\r_+)(1-2y_{-1}\r_-)} =
   \rt -y_{-1}\big[\r^2 +\rt^2\big] + O(y_{-1}^2)  \lb{rhoy}   \] 
from which we read off the infinitesimal transformations
$\d_{-1} \r = -2\r \rt$ and $\d_{-1} \rt = -\r^2 -\rt^2$. The last
two equations were already given in \c{Witt1} where they were derived 
by different methods. 

Let us observe that $\r(x,y_{-1})$ and $\rt(x,y_{-1})$
are new solutions to the original free field equations. This property
and the fact that this generalizes to the other fields had not been stressed 
before.  Note that the asymptotic flatness condition 
will be violated unless some care is exercised.
We may note also that the role played by the point $t=0$ in
previous treatments generalizes to the dilaton field $\r$ 
which is given by 
$\lim_{t\rightarrow 0}2w(x,t)t$. Let us emphasize, however,
that other points instead of $t=y=0$ in the domain of regularity could be
used to define the physical fields; this is the result of the action of the 
symmetry $s_{y_{-1}}$.

The formal similarity of \Ref{Mob2} and \Ref{L1trafo} is due to the
resemblance of two distinct subgroups of $\sl$, namely $\so$ and the subgroup
of upper triangular matrices, each subgroup serves however a different purpose.
Indeed one can rewrite \Ref{L1trafo} as
\[ s_y=v \, \pmatrix{ \cosh \! \th & \sinh \! \th \cr
              \sinh \! \th & \cosh \! \th \cr} (v_y)^{-1} \]
which is easily put with the help of \Ref{trigo} into the form of \Ref{Mob2}.

In fact there was another fortunate accident: one could identify the $\so$
subgroup of $\sl$  acting on $q$ with a subgroup of $\Ki$ acting on $t$ as we
have seen in \Ref{acc}\footnote{Admittedly,
we have not yet explained why $q(t)$ appears,
why $y$ is divalent or why the involution
$\I$ plays such a crucial role; we can only remark that it 
requires $u^2$ to appear as the fundamental variable in scalar equations.
Somehow the ``adjoint" dilaton representation involves $y$ or $w$ or $u^2$. 
The fundamental representation, however, will involve 
the variable $u$ or $t$ for the matter sector: this is one
way to rephrase the double valuedness of the mappings $t(q)$ and $t(y))$.}.
This fact will allow us to extend our group theoretical analysis of the 
dilaton sector to the full theory.

Let us now turn to the equations of motion. We have used them above
to define the dual potential $\rt$, but we could (and we did) forget about them
later in the discussion of the transformation group $\sl$.
For the usual non-linear $\s$-model 
$\r$ and $\rt$ in \Ref{v} would be independent fields even on shell; they
would then be subject to the second order non-linear field equations following
from the $\sl/\so$ $\s$-model Lagrangian (the analog of the second 
order equations \Ref{equP}) of which the free fields \Ref{rho} and
\Ref{rhotilde} are only very special solutions. By contrast in our case,
in order to recover the first order equation of motion
 $\p_\m \rt = - {\tilde \p}_\m \r$, 
we must actually impose a ``twisted" self-duality constraint 
\[ p_\m = \J \pt_\m  \;\;\; {\rm with} \;\;\; 
\J \equiv \L_{1}-\L_{-1} = \pmatrix{0&1\cr 1&0\cr} \lb{selfdual1}  \]
where $p_\m := \big( v^{-1}\p_\m v \big)^\perp $ with the 
superscript $^\perp$ denoting the projection onto the orthogonal
complement of the Lie algebra $\so$ in $\sl$. 
Since the matrix $\J$ is the generator $\so$ the twisted
self-duality condition \Ref{selfdual1} is obviously $\so$ invariant. 
This structure is one of the hallmarks of supergravity theories in even 
dimensions \c{Cremmer, Julia1, Julia3}. It is however a most complicated way
to bring free field theory into the framework of supergravity theories! In 
that form the $\sl$ invariance of the equations of motion becomes manifest 
as well as the local $\so$ invariance.
We shall verify in section 6 
that it is not an entirely artificial construction: it extends to all fields.

\section {The full dilaton coset space.} 
We would like to exhibit now the ``coset" structure associated with the full
$\Wir$ symmetry. Our aim here is to reinterpret and generalize the relation
\Ref{w} between the spectral parameters $y$ and $t$. 
The functional dependence \Ref{w} or \Ref{y} as functions of $t$ at fixed $x$ 
 did  emerge 
``on shell", i.e. after the imposition of \Ref{chi-t}, which implies the
dilaton equation of motion. Somehow on shell we only need the subgroup
$\sl$ but we now want to implement the $\Wir$ symmetry to go fully (i.e. on
a larger set of fields) 
off shell and to prepare for a simultaneous implementation of the 
$\Gi$ invariance.
For this purpose, we will replace the matrix $v$ of \Ref{v}
by another ``matrix" $\vh$ that will be the true analog of $\Vh$,
bearing the same relation to $\r$ as $\Vh$ to $\V$. As $v$ was in $\sl$,
$\vh$ should correspond to the exponentiation of the full $\Wir$ Lie 
algebra after gauge restoration. But as we explained in section 2, only
the ``regular" half of it can be exponentiated relatively easily. We shall
therefore be guided by the following formal (gauge-fixed) ansatz
\[ \vh_x  = \exp \sum_{n=0}^\infty \bigg( \chi_n (x) \,
  \L_n \bigg) \lb{expVir} \] 
that represents a map from $t$ to $y$ and where t should now be treated as an 
independent variable; we 
will thus exchange the status of the two variables $t$ and $y$ and forget their
qualificatives variable (resp. constant) as functions of $x$.

The operators $\L_n$ in the exponent may be chosen in order to
ensure invariance under the involution $\I$, so the application
\Ref{expVir} will  be a divalent map depending on $t$ through $q$. Note that
the condition for
a function of $t+\frac{1}{t}$ to be regular at $t=0$ is that it be regular as
a function of $q$ at $q=0$. 
As it stands \Ref{expVir} is only a formal expression,
but it is  obviously the infinite dimensional analog (pushed forward to $t$ 
space) of the triangular gauge used in \Ref{v}, where 
\[ v (\r , \rt)= \exp (\rt \, \L_1) \; \exp (-\log \r \, \L_0) 
      \lb{expVir1}      \]
While it was possible to represent \Ref{expVir1} as a two-by-two matrix, 
such a simple explicit formula is no longer available if the higher
generators $\L_m$ are taken into account, and this is
the reason why more work is required to make sense of \Ref{expVir}
beyond the purely formal level. 

We really want to generalize \Ref{w}
which is a relation between $t$ and $y$ and not only \Ref{yq} or \Ref{wq}
that relate $q$ and $y$; this is required by the fact that $t$ itself appears
in the matter sector. The existence of the involution $\I$ however 
suggests to keep the same $x$ independent relation \Ref{q} between $t$ and $q$.
We would also like to extend the transformation formula
\Ref{L1trafo} as well as the mapping \Ref{Mob2} or \Ref{coprod}.
We thus wish to interpret \Ref{expVir} as a representative element
of a certain infinite-dimensional cosetlike space which we will shamelessly
designate by $\Wir/\Ki$. Guided by previous experience with the 
coset space $\Gi/\Hi$, we are led to tentatively define this space 
as the set of (equivalence classes of) $x$-dependent maps 
$\vh_x(t):t\rightarrow \vh_x (t)\equiv Y(x,t)$, which are real analytic 
and (locally) invertible in some real neighborhoods of $t=0$ and $t=\infty$
and satisfy $Y(x,t)= Y(x,1/t)$\footnote{The analyticity domain should, 
however, be big enough to allow for the formulation and 
solution of a well-posed Riemann-Hilbert problem. 
It could depend on $x$. To be more precise, one should
regard $t$ and $Y$ as local coordinates on the Riemann sphere 
(or even a more general Riemann surface) and its double cover
with the two fixed points of $\I$ as the branch points.}. 
Due to the invariance 
under $\I$ the inverse map $t=t(x,Y)$ is double-valued over $Y$;
the two branches $t_\pm (x,Y)$ are related by 
\[ t_+(x,Y) = \frac{1}{t_-(x,Y)}  \lb{involution}  \] 
The above  condition
\Ref{involution} should also hold off shell. Like $\Vh$, $\vh$ depends 
on an infinite set of ``dual potentials" 
(the fields $\chi_n$ in \Ref{expVir}), 
and one must find suitable constraints so as to make $\vh$ 
depend on the single physical dilaton field $\r$ only; as in the 
matter sector this dependence will be non-local. 

Prior to the choice of gauge the global (rigid) semigroup 
$\Wir$ acts on $Y$ from the left by the transformations 
\[ Y(x,t) \rightarrow f \big(Y(x,t)\big) \;\;\; {\rm with} 
\;\;\; f \in \cR \lb{Di1} \]
where $\cR$ is defined as the set of germs of real analytic 
and (locally) invertible maps. Only the ``regular" 
transformations with $f(0)=0$
leave invariant the normalization condition
\[Y(x,0)= Y(x,\infty) =0 \lb{trY} \]
 which we will adopt as a preliminary 
definition of the triangular gauge.
In this gauge, we have $Y(x,t)= c(x) t + O(t^2)$ 
for sufficiently small $t$ and some nonzero function $c(x)$. The precise 
mathematical definition of triangularity presumably involves univalence on the 
unit disc of the $t$ plane or on a more general fundamental region of the 
involution $\I$. This has been discussed in detail in the matter sector as we
mentioned 
in section 2. The origin of the $t$ plane does play a special role in the
present treatment because it is at this point that one recovers the physical
fields: we recall the limit $\lim_{t\rightarrow 0}2w(x,t)t= \r$. But 
eventually a
``singular" $Y\equiv \vh$ could be used in a general gauge.

The gauge group $\Ki$ acts independently from the right according to 
\[ Y(x,t)\rightarrow Y\big(x,k(x,t)\big)   \lb{Ki}   \]
where $k(x,t)$ obeys \Ref{k(t)} pointwise for each $x$.
Two functions $Y_1$ and $Y_2$ are said to be gauge equivalent 
if there exists a gauge transformation $k(x,\cdot) \in \Ki$ 
such that $Y_1(x,t)= Y_2\big(x,k(x,t)\big)$ for all $x$. 
Clearly such gauge transformations may affect the normalization
condition on $Y$ defining the triangular gauge because 
they will move $t$ and $1/t$ about simultaneously
preserving their inverse relation. At least formally at 
the Lie algebra level, for ``singular" transformations \Ref{Di1}
with $f(0)\neq 0$, each $Y(x,t)$ can be brought back 
into the triangular gauge \Ref{expVir} by means of a $\K$ compensator 
(of course, no compensating transformations are required for the 
regular maps). The combined transformation  
\[ Y(x,t) \rightarrow f \Big(Y\big(x, k(t;Y,f)\big)\Big) \lb{Di}  \]
will then preserve the triangular gauge condition \Ref{trY} on $Y(x,t)$.
In the remainder of this section, we will study \Ref{Di} at the 
infinitesimal level in detail and discuss the uniqueness 
of the compensating transformation.

After all this preparation, we are now ready 
for a more precise construction of $\vh$.
The need for two spectral parameters $t$ and $y$ to build 
$\Hi$ (on the $t$ side) and $\Gi$ (on the $y$ side)
and eq. \Ref{coprod} suggest to simply rewrite as a first step 
the results of the previous section 
in terms of the function $y(t)\equiv y(q(t))$: 
\[ \vh_x  \big(t \big) = y(t;\r,\rt) ,  \lb{vhoft}            \]
with the special function \Ref{yq}; in this way, the $x$-dependent
family of maps $\vh_x : t \rightarrow y$ is  parametrized
by the single field $\r$ and its (first order) dual $\rt$.
In order to derive a ``linear system", we compose this map with its
inverse at infinitesimally shifted argument. The resulting map
$t\rightarrow t'=\vh_x^{-1}\circ \vh_{x+\d x}(t)$ is 
an $x$-dependent diffeomorphism on $t$ near the identity diffeomorphism; 
in the limit 
$\d x^\mu \rightarrow 0$, we can therefore associate with it
an element of the tangent space at the identity in the set of diffeomorphisms 
of $t$, i.e. an element of the Lie algebra of vector fields in $t$.
As usual the Lie algebra element can be described by the matrix product
$\vh^{-1}  \p_\m \vh$ in any linear representation (acting on the left) and 
when acting on the right
by composition on functions of $t$ it is realized by the differential operator
$-\d t(t)\frac{\p}{\p t}$ where the minus sign is a consequence of the right
action on functions. In our case this leads to the equivalence 
\[\vh^{-1}  \p_\m \vh :=
 - \p_\m y \bigg|_t \, \frac{\p t}{\p y} \bigg|_x \, \frac{\p}{\p t} \equiv 
  \p_\m t \bigg|_y \, \frac{\p}{\p t} 
\lb{vhdvh} \]
where the right action is responsible for the middle factor of the second
expression and the final sign is $+$. 
Taking the clues from \Ref{LS-t} we are thus led to demand
\[ \vh^{-1}    \p_\m \vh
 &:=& \bigg( \frac{1+t^2}{1-t^2} \r^{-1}\p_\m \r 
+\frac{2t}{1-t^2} \r^{-1}{\tilde \p}_\m \r \bigg) t\frac{\p}{\p t}  \non
 &\equiv& \r^{-1}\p_\m \r \, t_\ast \Big( q \frac{\p}{\p q} \Big) +
   \r^{-1}{\tilde \p}_\m \r \, t_\ast \Big(q^2 \frac{\p}{\p q} \Big)
\lb{vhdvh1} \] 
or
\[ \vh^{-1} \p_\pm \vh := 
\r^{-1} \p_\pm \r u^{\pm 1} \, t\frac{\p}{\p t}
 \lb{LS-t2}\]
The second line in \Ref{vhdvh1} is just the rewriting in terms of homographic 
maps (pushed forward from the $q$ to the $t$ variable by \Ref{t(q)}) of the 
results of section 
4 where one must use \Ref{harmr}, \Ref{v} and the computation of the matrix 
product that is implied in \Ref{selfdual1}. 
It provides a nice consistency check on our signs.

The main advantage  of rewriting the original linear system
\Ref{LS-t} in this fancy way is that \Ref{vhdvh} and 
\Ref{LS-t2} make manifest the group theoretical meaning of \Ref{LS-t}, we have
set up a formalism to introduce off shell fields for the realization of
the full $\Wir_y$ invariance. 
At this stage we are off shell if we treat $\rt$ as an 
independent field but we have only imbedded the M\"obius matrix group into
$\Wir_q$. We are still dealing with the subgroup $\sl$. A very important
property of \Ref{vhdvh} is the following:
the r.h.s of \Ref{vhdvh1} is an element of the
subalgebra $\Lie{\Ki_t}$ just like the r.h.s. of \Ref{LS-matter}
belongs to $\Lie{H^\infty_t}$. But now this mysterious difference from the
usual structure of (super)gravity theories can be understood:
In \Ref{selfdual1} the $\sl$ part of the field is not in $\so$ but in its 
orthogonal complement as it should according to the silver rules, it becomes
part of $\Lie{\Ki_t}$ simply because of \Ref{t(q)}, cf. the end
of section 2. Namely the full $\Wir_q$ corresponds by the map $t(q)$
to a singular coefficient extension of $\Ki_t$. This will be elaborated on in 
the next section. 
{}From here on we have inverted the point of view from
$t=t(y)$ to $y=y(t)$, by $\p_\m$  we shall always 
mean $\p_\m\big|_t$: we deform the image keeping the source of $\vh$ fixed.

The compatibility condition of the Lax pair \Ref{LS-t2} is just 
the dilaton equation of motion. Namely, a short calculation reveals that
\[ \p_+ \Big( \vh^{-1} \p_- \vh \Big) - \p_- \Big( \vh^{-1}\p_+ \vh \Big)
  + \Big[ \vh^{-1} \p_+ \vh \, , \, \vh^{-1} \p_- \vh \Big] :=
 \r^{-1} \Box \r \, \frac{4t^2}{1-t^2} \frac{\p}{\p t} \lb{vhdvh2} \]
where the commutator is now a commutator of the differential 
operators appearing on the r.h.s of \Ref{vhdvh1}, and $t$ is to be 
treated as an independent variable, so it must not be differentiated.
Clearly, \Ref{vhdvh2}
vanishes only if $\Box \r =0$. By duality, the on-shell condition
\Ref{LS-t2} relates all the higher order dual potentials 
$\chi_n$ to the dilaton field $\r$ in the same way as \Ref{LS-matter}
relates the infinitely many dual potentials occurring in the matter sector
to the basic physical field $\V$ and at the same time forces   
$\r$ to obey its equation of motion.

However, \Ref{vhoft} cannot be the most general solution to \Ref{LS-t2}.
Namely, by analogy with the matter sector, 
we expect, on shell, the most general
$\vh$ to depend on an infinitude of (real) integration constants $a_j$
in addition to the basic dual pair of fields $\r$ and $\rt$. This is the 
second step of our justification of \Ref{expVir}. In fact, the
dependence of $\vh$ on these extra variables, which are absent 
in \Ref{vhoft}, can be
made completely explicit by noting that, together with \Ref{vhoft}, 
every locally holomorphic and univalent function of the form
\[ Y \big(t;\r,\rt;a_0, a_1, a_2, \dots \big) :=  
 \sum_{j=0}^\infty a_j  y(t;\r,\rt)^{j+1} \lb{vhoft1} \]
with $a_0 \neq 0$ also solves \Ref{LS-t2}. To avoid overcounting
we may absorb the coefficient $a_0$ into $\r$ and $\rt$ by a rescaling,
and eliminate $a_1$ by shifting $\rt$ (and $w$) and suitably redefining  
the constants $a_j$, so that \Ref{vhoft1} becomes 
\[ Y\big(t;\r,\rt; a_2,a_3,\dots \big) \equiv Y(y, a_j) = y(t;\r,\rt) 
\Big( 1 + \sum_{j=2}^\infty a_j y(t;\r,\rt)^j \Big) \lb{vhoft2} \]
The function \Ref{y} is recovered as a special case by 
setting $a_j =0$ for $j\geq 2$ in \Ref{vhoft2}. Observe also
that the function $Y$ in \Ref{vhoft2} corresponds to
\[ \vh_x = \exp \big( \sum_{m\geq 2} c_m \L_m \big)
 \exp \big(\rt \L_1\big) \exp\big(-\log \r \L_0 \big) \lb{vhoft3}  \]
with constants $c_m, m\geq 2$. This formula shows that the general
solution of \Ref{LS-t2} can be understood as the composition of
a field dependent $SL(2,\R)$ map and a general Witt transformation
involving the higher generators with constant coefficients. 

More generally
(i.e. off-shell) the above formula should be regarded as a rigorous  
definition of the formal expression \Ref{expVir}, with the 
mutually independent fields $\r (x)$ and $\rt (x) \equiv a_1(x)$ 
and an infinite tower of extra fields $\{a_2(x),a_3(x),\dots \}$ serving
as a set of independent local coordinates of the 
infinite dimensional coset space \Ref{vhoft2} off-shell.
It is then the on-shell condition \Ref{LS-t2} that
enforces constancy of the $a_j$'s and turns $\r$ and $\rt$
into a dual pair of free fields, such that
repeated dualization merely reproduces the fields $\r$ and $\rt$ 
up to the integration constants $a_j$. We shall henceforth keep the notation
$Y$ instead of $\vh$ when we will be in this general framework. 

As already mentioned we are still in a fixed (triangular) gauge, and
we are looking for the combined global action of $\Wir$ and $\Ki$ 
``from the left" and ``from the right", respectively,
on the ``coset space" $\Wir/\Ki$ according to \Ref{Di}.
We will now study this action at the Lie algebra level and 
determine the induced variations on the coset space coordinates.
Infinitesimally, the combined action of the two groups is given by
\[ \Delta_m Y := \d_m Y - \frac{\p Y}{\p t}\bigg|_x \d_m t  \lb{Deltay1} \]
whose first part is the $\Lie{\Wir}$ transformation
\[ \d_m Y := -Y^{m+1} \lb{Deltay2}  \]
and whose second  part is the associated compensating $\Lie{\Ki}$
transformation of $t$, its minus sign is conventional but reminds us that the
infinitesimal shift of $t$ appears with a minus sign in the right action.
Given \Ref{Deltay1}, we can compute
the induced variations of the off shell
fields for all $m \in {\bf Z}$ by expanding 
\[ \Delta_m Y &=& \frac{\p Y}{\p y} y^2 \bigg( \d_m \rt -
    \ft12 \Big( t + \frac{1}{t} \Big) \d_m \r \bigg) +
   \sum_{j=2}^\infty \d_m a_j \, y^{j+1} \non
 &=& \frac{\p Y}{\p y} \r^{-1} \Big( -y \d_m \r +
    y^2 ( \r \d_m \rt - \rt \d_m \r ) \Big) +
  \sum_{j=2}^\infty \d_m a_j \, y^{j+1}    \lb{induced}   \]
we just have to substitute \Ref{vhoft2} into the l.h.s. and match 
coefficients of $y^j$; in deriving \Ref{induced} we must 
use the formulas \Ref{w} and  the simple result:
\[ \frac{\p y}{\p \rt} = y^2 \;\; , \;\;
   \frac{\p y}{\p \r}  = -\ft12 \Big(t + \frac{1}{t}\Big) y^2 \;\; ,\;\;
   \frac{\p y}{\p t} =  \ft12 \r \frac{1-t^2}{t^2} y^2.  \lb{dydrho} \]
The necessary
calculations are straightforward, though a bit tedious, and
we therefore just give the leading terms below. It might be
useful to notice that in the last formula of \Ref{induced} all
explicit $t$'s have been eliminated by use of \Ref{w}.

We first consider the ``regular" transformations corresponding to 
$m\geq 0$. Because these preserve the gauge condition $Y\big|_{t=0}=0$
we must set generically $\d_m t =\d_m y =0$. Following the procedure outlined
above we obtain, for $m\geq 2$,
\[  \d_m \r  &=&  \d_m \rt = 0 ; \non
    \d_m a_j &=&  0 \;\; {\rm for} \;\; j<m   \;\; ; \;\;
     \d_m a_m = -1 \; , \;  \d_m a_{m+1}=0  \non
    \d_m a_{m+2}&=& -(m+1)a_2  \; , \;
    \d_m a_{m+3}=-(m+1) a_3  \non
    \d_m a_{m+4}&=& -(m+1)(a_4+ \ft12 ma_2^2 ) \;\; , \; \dots    \lb{dmrho} \]
For $m=1$ and $m=0$ we similarly deduce that
\[ &&\d_1 \r = 0 \;\; , \;\; \d_1 \rt =-1 \non
&& \d_1 a_2 =0 \;\; , \;\; \d_1 a_3 = a_2 \;\; , \;\; \d_1 a_4 = 2a_3 \; ,\non 
 &&  \d_1 a_j = (j-2) a_{j-1} - \sum_{k+l=j-1} a_k a_l \;\; 
   {\rm for} \;\; j\geq 5 \;\; , \dots \lb{d1rho2}  \] 
and
\[ \d_0 \r = \r \;\; , \;\; \d_0 \rt = \rt \;\; , \;\;
   \d_0 a_j = j a_j \;\;\; {\rm for \;\; all} \;\; j\geq 2  \lb{d0rho1}  \]
In particular we have recovered the variations
of $\r$ and $\rt$ under $\L_1$ and $\L_0$ given in the preceding section.
From the formulas \Ref{dmrho} it is
evident that, for $m\geq 2$, the transformation
$\Delta_m Y= \d_m Y$ at lowest non-trivial order simply 
shifts the $a_j$'s by constants for $j\geq m$ leaving
the $a_j$'s for $j<m$ unaffected so that we can represent its
action on $Y\big( t;\r,\rt; a_2, a_3,\dots \big)$ by the differential
operator $-\p/\p a_m + \dots$.
For $m=1$, it shifts the first order dual potential $\rt$
in agreement with the result of section 4. Let us also note that these
transformations ``cancel out" in \Ref{LS-t2}.
In summary, the upper triangular half of $\Wir$ has exactly the
same effect on $Y$ as the upper triangular half of $\Gi$ on $\Vh$
in that it shifts the free integration constants arising in the
definition of the dual potentials.

Next we turn to the ``singular" transformations in $\Wir$ corresponding
to non-positive values of $m$. For these, $\d_m Y$ either 
does not vanish or diverges at $Y=0$, so compensating 
$\d k_m\equiv - \d_m t \frac{\p}{\p t}$
transformations are necessary. In order to determine them we must
as usual invoke the notion of triangular gauge, replacing the formal
definition \Ref{expVir} by  our definition of 
$\Wir/\Ki$ above. In order to stay inside it one must  
require the full variation $\Delta_m Y$ to vanish at $t=0$
when expressed as a function of $t$. This definition
is in accord
with the corresponding one for $\Vh$, cf. the 
discussion after \Ref{Geroch1}. Hence we choose the compensating $\Ki$ 
transformation $\d_m t$ in such a way that 
\[ \Delta_m Y \bigg|_{t=0} = 0         \lb{triangular}    \]
Let us remark that this prescription is similar (although 
not identical) to the one followed in section~3.4 of the second of
references Ref.~\c{Kir}.
For $\L_{-1}$, i.e. $\d_{-1} Y = -1$, \Ref{triangular} yields after using
\Ref{y} and again \Ref{dydrho}: 
\[  \d k_{-1} := -\d_{-1} t \, \frac{\p}{\p t} =
   \ft12 \r \big( 1-t^2 \big)\frac{\p}{\p t} =
  \ft12 \r \Big( \cL_1 - \cL_{-1} \Big)  
   \equiv  \ft12 \r \, \K_1 \lb{L1t}   \]
Not unexpectedly, this is just the infinitesimal version 
of the compensating gauge transformation found in 
\Ref{L1trafo}\footnote{In the sense that an infinitesimal shift
$\d \th = \varepsilon \r$ of the rotation angle $\th$ corresponds to
$\d t = \varepsilon \ft12 \r (1-t^2)$ as can be checked 
by varying \Ref{u}, compare \Ref{acc}.}. 
Similarly, for $\d_{-2} Y=-Y^{-1}$, we get
\[ \d k_{-2} := - \d_{-2} t \, \frac{\p}{\p t} 
  &=& \bigg\{ -\ft32 \r \rt \Big(1-t^2 \Big)
   + \ft14 \r^2 \Big( \frac{1}{t}-t^3 \Big)\bigg\} \frac{\p}{\p t} \non
  &=& -\ft32 \r \rt \, \K_1 + \ft14 \r^2 \, \K_2  \lb{L2t}  \]
In passing we note that all strict $\Ki_t$ transformations are such that
\[  \d_m t \Big|_{t=\pm 1} = 0  , \lb{tdt}   \]
because the points $t=\pm 1$ are fixed points of $\I$.

The induced variations of the fields can again be determined from 
the general formula \Ref{induced}. For $m=-1$,
we obtain  
\[ \d_{-1} \r = -2\r \rt   \;\;\; , \;\;\;  
   \d_{-1} \rt = -\r^2 - \rt^2 + 3a_2 \lb{Lrho1} \]
together with
\[ \d_{-1} a_2 = 4a_3 \;\; , \;\; \d_{-1} a_3 = 5a_4- 9a_2^2 \;\; , \;\;
   \d_{-1} a_4 = 6a_5-12 a_2 a_3 \;\; , \; \dots  \lb{Lrho1a}  \]
\Ref{Lrho1} differs from the familiar result of section 4 by the last constant
term.

The on-shell constancy of the $a_j$'s is preserved by  the induced variations. 
This fact may be proved directly by means of \Ref{induced} when one 
notices that in terms of $y$ the vector fields of $\Lie{\Ki}$ are combinations
with $x$-dependent coefficients of the vectors: 
\[ y^{-n}(1+2\rt y +(\rt^2-\r^2) y^2)\frac{\p}{\p y}. \]
There are three steps: one must first cancel 
the singular and constant terms in the 
expansion in powers of $y$ introduced by $\d_m Y$, then one 
deduces the variations of $\r$ and $\rt$ from the terms linear and quadratic 
in $y$ and the rest involves only constant coefficients that determine the 
induced variations of the coefficients $a_j$.
When $m\leq -2$, we have to make use of the Laurent expansions (with $n\geq 1$)
\[ Y^{-n} = \sum_{j=-n}^\infty b^{(n)}_j y^j   \lb{Laurent} \]
In particular to take one more example,  $m=-2$, we find 
\[ \d_{-2} \r = \r^3 + 3 \r \rt^2  -4a_2 \r  \;\; , \;\;
   \d_{-2} \rt = \rt^3 + 3 \rt \r^2  -4a_2 \rt +5a_3  \lb{L2rho}     \]
together with
\[ \d_{-2} a_2 = 6a_4-13a_2^2 \;\; , \;\; 
   \d_{-2} a_3 = 7a_5-33 a_2 a_3 \;\; , \; \dots \lb{L2rhoa}  \]

The truncation to the $\sl$ sector corresponding to the 
previous section is effected by putting all $a_j=0$.
From the above formulas it is then immediately evident that this 
is a consistent truncation in the sense that the constants
are not ``excited" by the $\sl$ generators $\L_1, \L_0$ and $\L_{-1}$
and that $\r$ and $\rt$ transform among themselves.
The existence of this truncation is the principal difference
between this presentation of the dilaton-axion system and that of the 
matter sector, where an analogous truncation to a finite number of
dual potentials was not possible. Finally, it is not difficult to  
verify that the Witt algebra $[ \d_m , \d_n ] = (m-n) \d_{m+n}$ 
is indeed satisfied for the above transformation rules; in 
particular, there is no central term in the 
dilaton sector which would extend the Witt algebra to a 
Virasoro algebra. This result is a consequence of the fact 
that we used a subgroup $\Ki$ such that the compensating transformations 
are unambiguous.

An important fact, which is still not obvious, is the form invariance of
\Ref{LS-t2} under gauge transformations. We have already remarked 
upon the fact that the $a_j$'s
mix among themselves and can consistently 
be put equal to independent constants on shell.
It is also a fact that $\r$ and $\rt$ remain dual of each other (a conformal
notion). More generally the induced nonlinear and non-local variations of 
$\r$ and $\rt$ are compatible with the form of the
linear system \Ref{LS-t2} in the sense 
that they preserve the $t$-dependence on the r.h.s., so that
\[ \d_m \Big( \vh^{-1}\p_\m \vh \Big) &=&
  \p_\m \d k_m + \Big[ \vh^{-1} \p_\m \vh \, , \, \d k_m  \Big]  \non
  &=&  \bigg( \frac{1+t^2}{1-t^2} \d_m \big( \r^{-1} \p_\m \r \big) 
      + \frac{2t}{1-t^2} \d_m \big( \r^{-1} {\tilde \p}_\m \r \big)
     \bigg) t \frac{\p}{\p t}   \lb{varvhdvh}    \]     
where again $t$ is to be treated as an independent variable, 
the commutator is a commutator of differential operators in $t$ and we have 
used the equations of motion relating $\r$ and $\rt$ and ensuring that the 
$a_j$'s are constants. In other words we have now established a $\Lie{\Wir}$
global on-shell symmetry of our models and even defined off shell fields with
simple transformation laws.

\section{Synthesis with matter sector}
The task that remains is to fuse the coset spaces 
$\Wir/\Ki$ and $\Gi/\Hi$ and to work out the action of the
Witt algebra on the matter fields and the conformal factor.
For this purpose we introduce a generalized coset field in the
triangular gauge taking its values in the space \Ref{bigcoset}
\[ \Yh (x,t) := \Big( Y(x,t) \, , \, 
   \Vh (x,t) \, , \, \lt (x) \Big)    \lb{Yh}  \]
where the first entry is to be interpreted in the sense explained
in the preceding section. At first $\Yh$ is  ``off shell", 
with infinitely many independent potentials. These potentials 
will be put on shell by imposing a generalized linear system for $\Yh$
analogous to \Ref{LS-matter} and \Ref{LS-t2}, which relate them 
to the basic physical fields by duality. $\Yh$ has an inverse
which is double-valued just like $t$ itself (cf. \Ref{involution})
\[ \Yh_\pm^{-1}(x,Y) := \Big( t_\pm (x;Y) \, , \,
   \Vh^{-1} \big( x,t_\pm (x;Y)\big) \, , \, 
     \lt^{-1}(x) \Big) \lb{Yhinv} \]
With our present understanding the $\Yh$'s do not form a group; 
rather $\Yh$ should be regarded as a generalized vielbein (our
largest yet!) which is acted upon by different symmetry 
transformations from left and right as we shall
now explain. To study the general action of $\Wir \semi \Gi$
and $\Ki \semi \Hi$ on $\Yh$, it will again not be necessary to 
impose right away  the equations of motion, since equations \Ref{WGionYh}
and \Ref{KHionYh} below are also valid off-shell.
As already mentioned several times, the first ``group" acts on $\Yh$
as a set of global (rigid) transformation from the left according to
\[ \big( f,g,a \big) * \Yh(x,t)&:=& \non && \hspace{-8em}
 \bigg( f\big( Y(x,t)\big)\, , \, g\big(Y(x,t)\big) \cdot \Vh (x,t) 
  \, , \,  a \lt (x) e^{\O (g,\Vh)} \bigg)     \lb{WGionYh}    \]
with $(g(y),a)\in \Gi$ as in \Ref{group1} and $f\in \Wir$.
The composition law has the semidirect product structure
\[ \Big( f_1, g_1, a_1 \Big) * \Big( f_2, g_2, a_2 \Big) :=
  \bigg( f_1 \circ f_2 \, , \, (g_1 \circ f_2) \cdot g_2\, , \,
   a_1 a_2 e^{\O (g_1,g_2)} \bigg)    \lb{VirG}    \]
generalizing \Ref{group1},
where the dot $\cdot$ denotes ordinary matrix multiplication and
$\circ$ stands for composition of functions. The action of the compensating 
gauge subgroup $\Ki\semi \Hi$ on $\Yh$ from the right is similarly given by
\[ \Yh (x,t) * \big( k,h \big) := 
  \bigg( Y\big( x,k(x,t)\big)\, , \, \Vh \big( x,k(x,t)\big) \cdot h(x,t) 
\, , \,  \lt (x) e^{\O(\Vh , h)} \bigg)  \lb{KHionYh}  \]
with $k\in\Ki$ and $h\in\Hi$ and group composition 
\[ \Big( k_1 , h_1 \Big) * \Big( k_2 , h_2  \Big) :=
   \Big( k_1 \circ k_2 \, ,\,  (h_1 \circ k_2) \cdot h_2 \Big)
      \lb{kh}   \]
(as a ``group", this is a subgroup of \Ref{VirG}).
We emphasize once more that contrary to section 2, the spectral
parameter $t$ is to be regarded as an independent variable here, 
so that in all formulas below $\p_\m \equiv \p_\m |_t$ will be understood.
Observe also that $\Ki \semi \Hi$ has no central term (it is in fact
the maximal subalgebra of $\Wir \semi \Gi$ with this property), and that
in \Ref{WGionYh}, $\Wir$ only acts on the first entry of $\Yh$. When
we combine the global action of $\Wir$ with certain compensating 
rotations to maintain the generalized triangular gauge (to be
defined below), the matter fields in $\Vh$ and the conformal factor
will thus ``feel" 
the action of $\Wir$ only through the compensating rotations.

To recover the equations of motion (that is, to go on shell) 
and to reduce the number
of degrees of freedom to the physical ones, we impose an
enlarged linear system combining \Ref{LS-matter} and \Ref{LS-t2} 
\[  && \!\!\!\!\!\!\!\!\!\!  \Yh^{-1} \p_\pm \Yh =  \non
 &=& \Big( \vh^{-1} \circ \p_\pm \vh 
     \, ,\, \Vh^{-1}\p_\pm \Vh - \p_\pm Y \frac{\p t}{\p Y}
           \Vh^{-1} \frac{\p \Vh}{\p t} \, ,
 \, \lt^{-1} \p_\pm \lt - \O'(\Vh, \Vh^{-1}\p_\pm \Vh ) \Big)   \non
&:=&   \Big( \r^{-1} \p_\pm \r u^{\pm 1} t\frac{\p}{\p t}  
   \, , \, Q_\pm + u^{\pm 1} P_\pm\, ,\, 0\Big) 
    \in  \; \Lie{\Ki} \oplus \Lie{\Hi} ,  \lb{YdY}    \]
where $\O'$ is the mixed cocycle introduced in \c{BM}. 
A new feature to be noted here is that the second term 
in the middle entry on the r.h.s. is now due to the 
semi-direct product rather than the explicit $x$-dependence of $t$.
More specifically, when computing the logarithmic derivative
of $\Yh$ we must evaluate the product
\[ &&\!\!\!\!\!\!\!\! \Yh^{-1} (x,Y) *  \Yh(x',t)   =  \\ && \!\!\!\!\!\!
 \Big( t\big(x,Y(x',t)\big)\, , \, \Vh^{-1} \big(x, t(x,Y(x',t))\big)
\cdot \Vh(x',t) \, , \, 
\lt^{-1}(x) \lt(x') e^{\O(\Yh^{-1}(x),\Yh(x'))} \Big)   
             \nonumber     \lb{YY} \]
at non-coincident points $x\neq x'$ before differentiating
and taking the limit $x'\rightarrow x$; the extra term in the
middle entry of \Ref{YdY} then arises from the shifted argument
$t(x,Y(x',t))\neq t$. Furthermore, the central 
charge $c$ is not an element of the subalgebra $\Lie{\Ki} \oplus \Lie{\Hi}$, 
and consequently the third entry in \Ref{YdY} vanishes identically,
thereby giving rise to the first order equations \Ref{equconf}
for the conformal factor \c{BM}. The linear system \Ref{YdY} thus
gives rise to all equations of motion.

When checking the compatibility of \Ref{YdY}
it must be emphasized that the $\p_\m \r$ terms 
in \Ref{equP} do not arise from the action of $\p_\m$ on $t$ 
as in \Ref{LS-matter}. Rather, with $t$ as the independent variable,
they originate from commutators such as 
\[ && \!\!\!\! \Big[ \vh^{-1} \circ \p_+ \vh \, , \,  \Vh^{-1} \p_- \Vh 
  - \p_- Y \frac{\p t}{\p Y} \Vh^{-1} \frac{\p \Vh}{\p t} \Big] := \non &=&
   \bigg[ \r^{-1} \p_+ \r \, \frac{t(1-t)}{1+t} \frac{\p}{\p t} \, , \,
    Q_- + \frac{1+t}{1-t} P_-  \bigg]  =
  \r^{-1}\p_+ \r \, \frac{2t}{1-t^2} P_-   \lb{comm}    \]  
Thus we see that it is now the semi-direct product structure 
of $\Ki \semi \Hi$ which generates the extra $\p \r$ term 
in the equation of motion \Ref{equP} instead of
the $x$-dependence of $t$ as before in \Ref{LS-t}. 

We have already introduced the notion of triangularity 
for $\Vh$ and $Y$ separately; the relevant formal  conditions can
now be combined into a single one on $\Yh$:
\[ \Yh (x,t) \bigg|_{t=0} = {\rm regular}    \lb{triYh}  \]
This triangularity condition again can be discussed 
independently of whether $\Yh$ satisfies \Ref{YdY} or not. 
In the first case, the physical fields appear at zeroth order 
in an expansion of $\Yh$ about $t=0$,
and the higher order potentials, which can be extracted 
as the coefficients of higher powers in $t$, are related by duality to 
them, so that the induced transformations become non-local. 
For off-shell $\Yh$, all the higher potentials (such as $\r(x), \rt (x)$ 
and $a_j(x)$ of the previous section) are independent fields, 
and the induced transformations are
perfectly local in terms of them. In the remainder of this section we
shall concentrate on the (gauge fixed) global transformations plus compensating
transformations but remain off shell until we finally check the invariance of 
the full linear system \Ref{YdY}.

The infinitesimal action of $\Gi$ on $\Vh$ in the 
triangular gauge has already been presented
in \Ref{Geroch1}; further, $\Gi$ does not act on $Y(x,t)$ at all.
Likewise, we have already discussed the infinitesimal action
of $\Wir$ and $\Ki$ on $Y(x,t)$ in the last section. It remains 
to work out the action of $\Wir$ on the matter fields, making use
of the results of the  foregoing section.
By \Ref{WGionYh}, $\Vh$ is sensitive only
to $\Ki \semi \Hi$, whose infinitesimal action is given by 
\[ \Vh^{-1} \Delta_m \Vh =  -\d_m t \, \Vh^{-1} \frac{\p \Vh}{\p t}
      + \d h_m   \lb{WonVh}    \]
where $\d_m t$ is the compensating $\Ki$ transformation
determined from \Ref{triangular}. For $\Vh$ to remain 
triangular, we thus need both $\Ki$ and $\Hi$ compensators
as the second term on the r.h.s.
is necessary to remove the poles at $t=0$ introduced by the first. 
The induced variations of the physical fields are determined from
\[  \V^{-1} \d_m \V = \Vh^{-1} \Delta_m \Vh \bigg|_{t=0} , \lb{WonV}  \]
Of course, \Ref{WonV} makes sense only because, by definition, 
the variation $\Delta_m$ preserves the triangular gauge.

Because $\d_m t$ vanishes for $m\geq 0$, the physical fields 
are thus inert under the Witt transformations for $m\geq 0$ 
by \Ref{WonVh}. For negative $m$, on the other hand, we do get
nontrivial variations; for instance,
\[ \V^{-1} \d_{-1} \V = \ft12 \r \, \Vh^{-1} \frac{\p \Vh}{\p t}
   \bigg|_{t=0}    \lb{WonV1}   \]
and 
\[ \V^{-1} \d_{-2} \V = -\ft32 \r \rt \, \Vh^{-1} \frac{\p \Vh}{\p t}
   \bigg|_{t=0} + \ft14 \r^2 \frac{\p}{\p t} \bigg( \Vh^{-1}
  \frac{\p \Vh}{\p t} \bigg)\bigg|_{t=0}     \lb{WonV2} \]
where we made use of \Ref{L1t} and \Ref{L2t}. 
A little algebra shows that
\[ \d_{-1} Q_\pm = 0 \;\;\; , \;\;\;
   \d_{-1} P_\pm = \mp \r P_\pm \lb{WonPQ1} \]
and
\[ \d_{-2} Q_\pm + \d_{-2} P_\pm &=& \pm 3\r \rt P_\pm -
   \ft12 \r \p_\pm \rt \, \Vh^{-1} \frac{\p \Vh}{\p t} \bigg|_{t=0} \non
 &&   + \r^2 P_\pm + \ft12 \r^2 \bigg[ P_\pm \, , \, \Vh^{-1}
    \frac{\p \Vh}{\p t}\bigg|_{t=0} \bigg]  \lb{WonPQ2}  \]
from which $\d_{-2} Q_\m$ and $\d_{-2} P_\m$ immediately
follow by projection. Checking the commutators
$[ \d_m , \d_{-1} ]$ and $[ \d_m , \d_{-2} ]$, and recalling that
the full algebra can be generated from $\L_m$ for $-2\leq m \leq 2$, 
it is straightforward to verify the absence of a central term in 
the action of $\Wir$ on the matter fields.

The connection of the above results with those of \c{Witt1,Witt2}
is established by noting that the variations of one half
of the Witt algebra on the matter fields can be combined
into a single generating function according to
\[ \d (s) \Vh(x,t) := \sum_{m=1}^\infty s^m \Delta_{-m} \Vh (x,t)
     \lb{generate}  \]
in terms of a parameter $s$. What the method of generating functions
cannot capture, however, is the action of the full Witt algebra 
in the dilaton sector exhibited in the previous section. Besides,
\Ref{generate} makes sense only in the triangular gauge.

The action of the symmetries
 on the central term can also be analyzed.
{}From \Ref{WGionYh} and \Ref{KHionYh} together with the formulas 
given in the appendix of \c{BM} we deduce that, quite generally,  
\[ \d \s =  \O'(\d g , \Vh) +
   \O' \Big( \Vh , \d h \Big)   \lb{Wonsigma}  \]
where $\O'$ is again the mixed cocycle. Therefore, and by \Ref{WonVh},
$\lt$ ``sees" the action of $\Wir$ only via $\Hi$; substituting the 
formulas of the foregoing section and taking into account
\Ref{WonVh} we arrive at
\[ \d_m \s &=& \ft12 \oint_{t=0} \frac{dt}{2\pi i}\,
   {\rm Tr} \Big( \Vh^{-1} \frac{\p \Vh}{\p t} \d h_m \Big) \non
  &=&  \ft12 \oint_{t=0} \frac{dt}{2\pi i}\,  \d_m t \,
     {\rm Tr} \bigg( \Vh^{-1} \frac{\p \Vh}{\p t} \bigg)^2 \lb{Wons} \]
It is obvious from these results that $\d_m \s =0$ for $m\geq -1$, for
which $\d_m t$ has no poles (or vanishes), and because the logaritmic
derivative of $\Vh$ is regular for triangular $\Vh$.
For $m=-2$ we obtain
\[ \d_{-2} \s = -\ft12 \r^2 \, {\rm Tr} \bigg( \Vh^{-1}\frac{\p \Vh}{\p t} 
  \Big|_{t=0}  \bigg)^2   \lb{Wons2}  \]
We have tested the general formula \Ref{Wons} by substituting
\Ref{WonPQ2} and \Ref{L2rho} into the equation of motion
for the conformal factor \Ref{equconf}. The commutator algebra
on $\s$ again takes the form of the Witt algebra:
\[ [ \d_m , \d_n] \s = (m-n) \d_{m+n} \s \lb{commsigma}, \]
In summary, there is no central term on any of the fields
in the induced algebra. However, in view of the possible 
extension of the affine to a hyperbolic Kac Moody symmetry, 
we would like to point out that the absence of a central 
term may simply be due to the fact that the matter fields,
the dilaton and the conformal factor realize only the level zero 
sector of the theory. Thus a central term may yet appear at
non-zero level, just like for the Sugawara construction, but this will 
require finding the corresponding higher level dual potentials.

Finally, we can construct conserved currents for $\Wir$ and $\Gi$; 
in accordance with the silver rules they both live on the $y$ side.
For $\Wir$, we take
\[ j_\m &=&  \p_\m \vh \, \vh^{-1} := 
  \p_\m t \bigg|_Y \frac{\p Y}{\p t}\bigg|_x \frac{\p}{\p Y} \non
 &=& - \p_\mu Y \Big|_t  \frac{\p}{\p Y}  
\lb{current}\]
Since the equation of motion forcing the ``constant spectral parameter" 
to be really constant is $\p_\m Y =0$, this current is not 
only conserved but actually vanishes on shell.
The corresponding conserved Kac Moody current was already given
in \c{Nic}. Using the results from section 3,
it can be cast into the form 
\[ \Jt_\m = \r (y) \Vh \Big( \cosh \! \th P_\m - 
             \sinh \! \th \Pt_\m \Big) \Vh^{-1}  
          \lb{current1}     \]
with $\r (y)$ from \Ref{rhoy}. This current can be viewed as 
the $\L_{-1}$ transform of the corresponding conserved current 
associated with the underlying finite dimensional group $G$ 
\[ \J = \r \V P_\m \V^{-1}      \lb{Gcurrent}      \]
with the spectral parameter $y$ serving as the $\L_{-1}$ transformation
parameter. Taking into account the (constant) prefactor $w^2$ 
by which \Ref{Gcurrent} differs from the expression given in \c{Nic}, 
one reobtains the elegant formula given there:
\[ \Jt_\m = {\tilde \p}_\m \bigg( \frac{\p \Vh}{\p y} \Vh^{-1} \bigg)
             \lb{current2}     \]

The semi-direct product structure $\Wir \, \semi \, \Gi$ enables us to
understand the   
matter linear system \Ref{LS-matter} as an $e^{y\L_{-1}}$ transform
of the ordinary equation of motion \Ref{VdV}, with 
$y_1:=y$ the constant spectral parameter.
In view of the existence of similar parameters $y_m$ associated with the
generators $\L_{-m}$ for $m\geq 2$, this suggests the possibility
that the full theory may actually require infinitely many
spectral parameters for its manifestly symmetric  formulation: a hierarchical
structure. The higher
scattering parameters should be related to the integration constants of the
higher dual dilatonic potentials mentioned above, in the same way as the
inverse of $y$ is related to the integration constant that appears in the
definition of  the first dual potential $\rt$ \Ref{w}. \Ref{selfdual1} 
begs for a generalization and there are indications that it exists.
Finally further mathematical work is also 
needed to bring in more analysis and enrich our
present understanding of the ``group" $\Wir$.

\section {An action and symplectic structure}
An important question concerns the existence of a  
symplectic realization of the infinite dimensional symmetries
discussed in this paper. Even ignoring certain technical difficulties
caused by non-ultrolocal terms in the Poisson brackets,
there are numerous indications that, at least for the flat space
integrable systems, the canonical brackets of the conserved charges 
do not reproduce the affine algebra found at the level 
of the equations but rather a Yang Baxter type structure 
for the classical theory \c{PL, Davies, Korep}.
As a contribution to the study of this question we present an 
alternative action for the truncated dilaton sector and the  
equations of motion given in section 3 and 4.

The covariant Lagrangian is\footnote{For Euclidean worldsheets,
the ``$\th$-term" involving $\rt$ must be multiplied by a factor of $i$.}
\[ {\cal L} =\te^{\m \n}\o_\m \p_\n \r  + \gt^{\m \n} \o_\m \p_\n \rt 
 +  \ft12 \r \gt^{\m \n} \, {\rm Tr} \, P_\m P_\n \lb{action} \] 
with $\te^{\m \n}:= e e_\a^{\;\;\m} e_\b^{\;\; \n} \ve^{\a \b}$ and
$\gt^{\m \n}:= e e_\a^{\;\; \m} e_\b^{\;\; \n} \eta^{\a \b}$, 
where $e_\m^{\;\;\a}$
is the zweibein, $e_\a^{\;\; \m}$ its inverse, and $e$ its determinant.
The main advantage of this action is that the fields
$\r$ and $\rt$ are treated in a more symmetric fashion, and that it 
separates the conformal factor from the unimodular gravitational degrees
of freedom in a covariant way, since \Ref{action} is manifestly 
invariant with respect to $2d$ diffeomorphisms and,
unlike the original action, under Weyl transformations
as it depends only on $\gt^{\m \n}$. It is now easy to check 
that \Ref{action} gives rise to all the equations of motion. 
Varying $\o_\m$, we get
\[ \te^{\m \n} \p_\n \r + \gt^{\m \n} \p_\n \rt = 0 \lb{covrho}  \]
which is just the the covariant version of the self-duality constraint. 
Normally, it would be illegal to
impose the selfduality constraint by a Lagrange multiplier because
this would introduce new propagating degrees of freedom.
What saves the
day here is that the Lagrange multiplier $\o_\m$ is related 
to a field already present in the model, namely the conformal factor.
To see this we first vary $\rt$ to deduce 
$\o_\m = \te_\m^{\;\; \n} \p_\n \s$ with some scalar field $\s$.
Now varying the unimodular metric $\gt^{\m \n}$, we get (the 
covariant analog) of the first order equations \Ref{equconf}.
This permits us to identify the field $\s$ with the one
introduced in \Ref{loglt}. Finally,
varying $\r$ we get the second order equation
$\p_\m (\gt^{\m \n} \p_\n \s ) = \ft12 \gt^{\m \n}\, {\rm Tr} \, P_\m P_\n $.
which, however, can be shown to follow from the other
equations given in section 1, and is therefore not independent.

From the results of the previous two sections it follows that
all the equations of motion are invariant under a non-linear
and non-local realization of the Witt algebra, unlike the
action \Ref{action}. What is important
here is that we must distinguish the Witt symmetries acting 
on the {\em field space} spanned by $\r$ and $\rt$ together with 
the infinitely many constants $a_j$ from the Weyl transformations
and the space-time symmetries (diffeomorphisms on the worldsheet) 
which are manifest in \Ref{action}. It is only in special gauges
(such as Weyl canonical coordinates for axisymmetric stationary
solutions) that these symmetries are tied together
and the Witt symmetry acting on the fields becomes a symmetry
acting on the space-time coordinates.

The canonical momenta for the dilaton-axion system following
from \Ref{action} are readily computed. Assuming the conformal
gauge for simplicity, i.e. $\gt^{\m \n} = \eta^{\m \n}$, we obtain
\[  \Pi := \frac{\d \cL}{\d (\p_0 \r )} = \o_1 = - \p_0 \s
\;\;\; , \;\;\; \tilde \Pi 
:= \frac{\d \cL}{\d (\p_0 \rt )} = \o_0 = -\p_1 \s \lb{momenta1} \]
On the other hand, the canonical momenta associated with $\o_\m$ vanish
\[ \O^\mu := \frac{\d \cL}{\d (\p_0 \o_\m)} = 0  \lb{momen2}   \]
Equations \Ref{momenta1} and \Ref{momen2} are second class constraints.
The secondary constraint $\p_0 \O^\mu \propto \{ \cL , \O^\mu \}=0$ 
implies the equations of motion \Ref{covrho}. Defining
\[ \Pi_\pm := \Pi \pm \tilde \Pi = \pm \o_\pm = - \p_\pm \s
     \lb{momenta2}       \]
we get the equal time Dirac brackets
\[ \Big\{ \Pi_+ (x) \, , \, \r_+ (x') \Big\} =
   \Big\{ \Pi_- (x) \, , \, \r_- (x') \Big\} =
   \d (x - x')    \lb{bracket}   \]
where $x,x'$ are spatial coordinates,
with all other equal time brackets equal to zero. Adapting these
formulas to the Euclidean case one arrives at the canonical brackets
already postulated in \c{KN1}.

\noindent
{\bf Acknowledgments.} We would like to thank P.~Breitenlohner, 
J.L.~Gervais, D.~Korotkin, D.~Maison and P.~Slodowy for discussions related 
to this work. H.~N. is especially grateful to P.~Breitenlohner and
D.~Maison for numerous conversations over the years; his work was
supported by the EU through the contracts
ERBCHRXCT920069 and SCI CT91-0650. 
B.~J. would like to thank the II. Institute of Theoretical Physics
for hospitality and the Alexander von Humboldt Foundation for a very 
stimulating invitation.

\end{document}